\documentclass[11pt,a4paper]{article}
\usepackage[a4paper,left=2.5cm,right=2.5cm,top=3cm,bottom=3cm]{geometry}

\usepackage{graphicx}
\usepackage{multirow}%
\usepackage{amsmath,amssymb,amsfonts}%
\usepackage{amsthm}%
\usepackage{mathrsfs}%
\usepackage[title]{appendix}%
\usepackage{xcolor}%
\usepackage{textcomp}%
\usepackage{manyfoot}%
\usepackage{booktabs}%
\usepackage{algorithm}%
\usepackage{algorithmicx}%
\usepackage{algpseudocode}%
\usepackage{listings}%
\usepackage{float}
\usepackage{comment}
\usepackage{mathtools}
\usepackage[colorlinks=true, allcolors=blue]{hyperref}

\title{Edge-based mean-field approximation of dynamics on networks via approximate lumping of Markov chains}
\author{G\'abor Tim\'ar$^{1,*}$, Jonathan A. Ward$^1$ and P\'eter L. Simon$^2$\\[0.5em]
\small $^1$School of Mathematics, University of Leeds, Leeds, UK\\
\small $^2$Institute of Mathematics, E\"otv\"os Lor\'and University, Budapest, Hungary\\
\small $^*$Author to whom any correspondence should be addressed.\\
\small \texttt{g.timar@leeds.ac.uk}}
\date{}

\begin{document}

\maketitle

\begin{abstract}
\noindent
Mean-field approximations for dynamical processes on networks are widely used, but existing derivations often rely either on moment closures or on idealised assumptions about network structure, leaving the nature of the underlying averaging unclear.
Here we present a mathematically principled framework for deriving edge-based mean-field approximations for a broad class of Markov processes on networks using approximate lumping. We consider models in which each vertex is in one of a finite number of vertex states and transitions depend on the number of neighbours in each state. Our approach partitions the full Markov chain state space according to the number of vertices and edges in each possible state, and averages transition rates between partitions. This yields density-dependent population processes that, in the limit of large system size, reduce to a low-dimensional system of ordinary differential equations. We demonstrate the method on single graphs and graph ensembles, such as Erd\H{o}s-R\'enyi random networks, and show that well-known edge-based mean-field approximations arise as special cases of our approach. Our approximate lumping framework clarifies the nature of the averaging underlying mean-field approximations, providing a basis for future work on assessing their accuracy.
\end{abstract}

\vspace{0.5em}

\vspace{1em}

\section{Introduction}
\label{sec1}

The behaviour of many real-world complex systems can be described as a continuous-time Markov process on the given underlying network of interactions \cite{barrat2008dynamical, porter2016dynamical, kiss2017book, newman2018networks}. Such dynamical systems include, for example, spin systems \cite{dorogovtsev2002ising, leone2002ferromagnetic, dorogovtsev2004potts, dorogovtsev2008critical}, opinion dynamics
\cite{galam2002minority, sood2005voter, sznajd2000opinion} and the spread of information or epidemics within a population \cite{pastor2015epidemic, kiss2017book}.
Due to the complexity of these processes, mean-field approximations are often used in their analysis \cite{gleeson2011high, fennell2019multistate, kiss2017book}, which neglect aspects of the network structure \cite{gleeson2012accuracy} and/or dynamical correlations between the states of neighbouring vertices.

Mean-field descriptions of dynamical processes on networks are typically derived in one of two broad ways. A common approach is to begin from an exact but unclosed system of equations for vertex-level and higher-order joint probabilities, and then impose a closure approximation at a chosen level of the hierarchy, for example by approximating triples in terms of pairs or pairs in terms of single-vertex marginals. This strategy underlies classical pairwise and higher-order models in network epidemiology and interacting particle systems \cite{keeling1999effects, sharkey2011deterministic, house2011insights, taylor2012markovian, rand1999correlation}. An alternative approach is to derive equations that are exact descriptions for certain idealised network structures (for example, no degree-degree correlations or no finite loops) and then interpret these equations as approximations for more general networks \cite{karrer2010message, miller2012edge, sharkey2015exact}.
In both approaches, bounds on the true dynamics are available in certain settings \cite{karrer2010message, vanmieghem2014exact, wilkinson2017relationships, simon2018bounding} but in general it is difficult to assess how good the approximations can be expected to be for a given real-world network structure. Furthermore, even the question of what precisely is being averaged in these mean-field approximations is unclear.

We consider a mathematically principled approach for deriving mean-field approximations for a wide range of dynamical models on networks, starting from the exact Markov chain descriptions of the given
dynamics. Specifically, we consider models in which each vertex of a network is in one of a finite number of vertex states at any given time, and transition rates between the vertex states of a given vertex
depend only on the vertex states of its neighbours. We apply a concept called approximate lumping \cite{kemeny1960finite, buchholz1994exact, ward2022micro, ward2025mean}, where states of the Markov chain are grouped together according to a predefined partition, and transition rates between states that lie in distinct partitions are averaged. This scheme lends itself naturally to partitions of the state space where all states within a partition have the same number of vertices in each of the possible vertex states \cite{ward2022micro}. This results in density-dependent population models. In the limit of large system size, the description of such models can be reduced to a small system of ordinary differential equations \cite{kijima1997markov}.
A recent work \cite{ward2025mean} refined this approach by considering arbitrary vertex partitions on a network, defining lumping partitions as sets of states that, for each vertex partition, have the same number of vertices in each vertex state. It was shown that, using approximate lumping of this kind, higher-resolution mean-field approximations, such as degree-based \cite{pastor2001epidemic, pastor2001bepidemic, moreno2002epidemic} and individual-based mean-field \cite{van2008virus, sharkey2008deterministic, youssef2011individual}, may be derived.

In this paper we build on the work in Refs. \cite{ward2022micro, ward2025mean} and apply the approximate lumping approach to derive more accurate, edge-based mean-field (EBMF) approximations.
This represents a refinement of the vertex-count-based population processes used in Refs. \cite{ward2022micro, ward2025mean}, by incorporating edge-state information. More precisely, we use a lumping scheme in which all states within a lumping partition share not only the same numbers of vertices in each possible vertex state, but also the same numbers of edges in each possible edge state. The resulting edge-based population processes are density-dependent \cite{ethier2009markov, ward2025mean} and provide approximations to the original dynamics. In the large-network limit, these processes become deterministic and are governed by a low-dimensional system of ordinary differential equations. We demonstrate how these sets of ODEs can be obtained for dynamics on single graphs as well as ensembles of graphs. We show that well-known EBMF approximations \cite{kiss2017book} are special cases of our edge-based approximate lumping approach, thereby placing them within a principled mathematical framework.

The paper is organised as follows. In Section \ref{sec2} we introduce the necessary mathematical background, including the continuous-time Markov chain framework, the concept of lumping and density-dependent processes. Section \ref{sec3} presents the general formulation of edge-based lumping, illustrated through the example of binary-state dynamics. In Section \ref{sec4} we apply the approach to a single graph, using a cycle as a case study. Section \ref{sec5} extends the framework to graph ensembles, specifically random regular and Erd\H{o}s-R\'{e}nyi networks. In Section \ref{sec6} we conclude with a discussion of the results.

\section{Mathematical background}
\label{sec2}

In this section we introduce the class of models studied in the paper, together with the concepts of Markov chain lumping and density-dependent processes. Since this model class is the same as the one considered in our previous work \cite{ward2025mean}, parts of the present section closely follow the exposition given there.

\subsection{Model setup: dynamical processes on networks as continuous-time Markov chains}
\label{sec21}

Let $G=(V,E)$ denote a graph with vertex set $V$ and edge set $E$, where the number of vertices is $N = |V|$ and the number of edges is $L = |E|$. Unless otherwise stated, we consider dynamical processes on finite connected simple networks (i.e. undirected, unweighted, with no self-loops or multiple edges) described by continuous-time Markov chains where each vertex can be in one of a finite number $M$ of \emph{vertex states} and the set of possible vertex states is $\mathcal{W}=\{\mathcal{W}_1,\mathcal{W}_2,\dots,\mathcal{W}_M\}$.
(The restriction to simple networks is made for notational clarity. Our construction extends directly to networks with multiple edges by treating them as distinct incident contacts. Self-loops would require a separate convention and will not be considered here.)
The state space $\Omega$ of the Markov chain is the set of all vertex state configurations: $M^N$ states in total. If the graph is in state $S\in \Omega$ then the vertex state of vertex $v\in V$ will be denoted by $S(v)$. From here onwards we will refer to the states $S\in \Omega$ as \emph{microstates}, to clearly distinguish them from vertex states and \emph{lumped states}, which will be introduced in Section~\ref{sec22}. The number of microstates is extremely large for even moderate $N$, however, since $\Omega$ is finite we can enumerate the microstates so that $\Omega=\{S^{[1]},S^{[2]},\dots,S^{[M^N]}\}$.

We assume that the dynamics are governed by local, homogeneous, single-vertex transition (SVT) models, where a vertex changes vertex state at a rate that is a function of only the number of its neighbours in each vertex state, and the rate function is the same for all vertices. Note that a more general class of models would allow for behaviours in which multiple vertices change vertex states at once, for example if a vertex exports its vertex state to its neighbours \cite{ward2018general}. SVT models can be represented by the pair $(\mathcal{W},\mathbf{R})$, where $\mathbf{R}$ is a matrix\footnote{Throughout, both vector and matrix quantities will be written in bold.} of functions that dictate the rates at which vertices change vertex state. More precisely, for each $\mathcal{A},\mathcal{B}\in \mathcal{W}$, we associate a function $R_{\mathcal{A},\mathcal{B}}:\mathbb{Z}^M_{\ge 0}\rightarrow\mathbb{R}_{\ge0}$, where $R_{\mathcal{A},\mathcal{B}}(d_1,d_2,\dots,d_{M})\ge0$ gives the rate that a vertex in vertex state $\mathcal{A}$ changes to vertex state $\mathcal{B}$ if it has $d_1$ neighbours in vertex state $\mathcal{W}_1$, $d_2$ neighbours in vertex state $\mathcal{W}_2$, etc. If transitions between a pair of vertex states $\mathcal{A},\mathcal{B}\in \mathcal{W}$ do not occur in a particular model, then the rate $R_{\mathcal{A},\mathcal{B}}$ is identically zero. We may think of the model $\mathcal{M}=(\mathcal{W},\mathbf{R})$ as a directed graph over vertex states where a directed edge goes from vertex state $\mathcal{A}$ to $\mathcal{B}$ if $R_{\mathcal{A},\mathcal{B}}$ is not identically zero.
We call $\mathbf{R}$ the vertex state transition matrix (VSTM). In this paper, as in Ref. \cite{ward2025mean}, our main focus will be on VSTMs that are affine functions, so

\begin{align}
R_{\mathcal{A},\mathcal{B}}(d_1,d_2,\dots,d_{M})=\zeta_0^{\mathcal{A},\mathcal{B}}+\sum_{m=1}^{M}d_m \zeta_m^{\mathcal{A},\mathcal{B}},
\nonumber
\end{align}

\noindent
where all of the constants $\zeta_m^{\mathcal{A},\mathcal{B}}$ are non-negative. Most SVTs have VSTMs of this form \cite{ward2019exact}, although notable exceptions include the non-zero temperature Ising-Glauber dynamics \cite{glauber1963time}, the nonlinear $q$-voter model \cite{castellano2009nonlinear} and threshold models \cite{watts2002simple}.

Given the graph $G$ and model $\mathcal{M}$, we construct the corresponding continuous-time Markov chain. Let $\mathbf{P}(t)=(P_1(t),P_2(t),\dots,P_{M^N}(t))^{\rm T}$ be the time-dependent Markov chain probability distribution over $\Omega$, where $P_i(t)$ is the probability of being in microstate $S^{[i]}$ at time $t$. The evolution of $\mathbf{P}(t)$ is then given by the forward Kolmogorov or master equation \cite{kijima1997markov},

\begin{align}
\dot{\mathbf{P}}=\mathbf{Q}^{\rm T} \mathbf{P},
\label{eq:21.20}
\end{align}

\noindent
where $\mathbf{Q}$ is the infinitesimal generator, an $M^N \times M^N$ matrix in which each off-diagonal component $Q_{kl}$ gives the transition rate from $S^{[k]}$ to $S^{[l]}$, and the diagonal components ensure that rows sum to zero so that probability is conserved. We assume that a vertex changes vertex state instantaneously, so transitions only occur between pairs of microstates that differ in exactly one vertex state. We call such pairs of microstates \emph{transition pairs} and
use the notation $S^{[k]}\stackrel{v}{\sim}S^{[l]}$ to indicate that the microstates $S^{[k]}$ and $S^{[l]}$ form a transition pair with \emph{transition vertex} $v$, i.e., if $S^{[k]}\stackrel{v}{\sim}S^{[l]}$ then $S^{[k]}(v)\ne S^{[l]}(v)$ and $S^{[k]}(u)=S^{[l]}(u)$ for all $u\ne v$.
For vertex $v$ and microstate $S^{[k]}$ let 
$d^{[k]}_m(v)$ be the number of neighbours of $v$ with vertex state $W_m$ in microstate $S^{[k]}$.
Thus the transition rate between microstates $S^{[k]}$ and $S^{[l]}$ in homogeneous SVT models is given by

\begin{align}
Q_{kl}=\left\{\begin{array}{cc} R_{S^{[k]}(v),S^{[l]}(v)}(d^{[k]}_1(v),d^{[k]}_2(v),\dots,d^{[k]}_{M}(v))
& \text{if $S^{[k]}\stackrel{v}{\sim}S^{[l]}$}\\ 0 &
\text{otherwise}
\end{array}\right..
\nonumber
\end{align}

\subsection{Coarse-graining via lumping}
\label{sec22}

We consider \emph{lumping} of Markov chains \cite{kemeny1960finite}, a form of coarse-graining.
For a Markov chain described in Sec. \ref{sec21} let us partition the microstate space according to a \emph{lumping partition} $\Pi_\Omega=\{\Omega_1,\Omega_2,\dots,\Omega_{n}\}$.
When the system is in a microstate $S^{[k]} \in \Omega_i$ we say that it is in the \emph{lumped state} $\Omega_i$. (We use the same notation for lumping partition cells and lumped states. We will always clarify which one we mean, whenever the distinction is important and not obvious from context.)
For a given lumping partition, it will be useful to introduce the \emph{collector matrix} $\mathbf{C}\in\{0,1\}^{M^N\times n}$ (see Ref. \cite{buchholz1994exact}), whose $kj$th component is

\begin{align}
C_{kj}=\left\{\begin{array}{cc}
    1 & \text{if $S^{[k]}\in \Omega_j$},\\
    0 & \text{otherwise}. \end{array}\right.
\nonumber
\end{align}

\noindent
Let us define an $n$-dimensional Markov chain by introducing an $n \times n$ generator matrix $\mathbf{q}$ for a vector $\mathbf{p}(t)=(p_1(t),\dots,p_n(t))^T$, the time-dependent Markov chain probability distribution over $n$ states. The evolution of $\mathbf{p}(t)$ is then determined by the master equation

\begin{align}
\dot{\mathbf{p}}=\mathbf{q}^{T} \mathbf{p}.
\label{eq:22.20}
\end{align}

\noindent
If $\mathbf{q}$ can be defined such that $\mathbf{p}(t) = \mathbf{C}^T \mathbf{P}(t)$ for all $t$ (i.e. that $p_i(t)$ is the sum of microstate probabilities over all microstates in lumped state $\Omega_i$ for all time), then the lumping partition $\Pi_\Omega$ is called an \emph{exact lumping}. If no such $\mathbf{q}$ exists, then $\Pi_\Omega$ is called an \emph{approximate lumping}. A necessary and sufficient condition for a lumping to be exact is that the sum of transition rates out of a microstate $S^{[k]}\in \Omega_i$ into the cell $\Omega_j$ is the same for all microstates in the cell $\Omega_i$. This condition is equivalent to the existence of a matrix $\mathbf{q}$ such that

\begin{align}
\mathbf{QC}=\mathbf{Cq},
\label{eq:22.30}
\end{align}

\noindent
which we call the \emph{lumpability condition}.

For an exact lumping, the matrix $\mathbf{q}$ is unique and may be explicitly determined by considering any left inverse of the collector matrix, i.e., any matrix $\mathbf{D}\in\mathbb{R}^{n\times M^N}$, such that $\mathbf{DC} = \mathbf{I}$, the identity matrix. Multiplying Eq. (\ref{eq:22.30}) by such a $\mathbf{D}$ matrix we get

\begin{align}
\mathbf{q}=\mathbf{DQC},
\label{eq:22.40}
\end{align}

\noindent
which is unique as long as the lumping associated with matrix $\mathbf{C}$ is exact.

When the lumping is not exact, the generator matrix $\mathbf{q}$ from Eq. (\ref{eq:22.40}), using any $\mathbf{D}$ matrix, will produce a function $\mathbf{p}(t)$ which is not equal to $\mathbf{C}^T \mathbf{P}(t)$ for all $t$, but is instead an approximation. [We assume that the initial conditions can still be chosen such that $\mathbf{p}(0) = \mathbf{C}^T \mathbf{P}(0)$, but the two trajectories will diverge for $t > 0$.] In this case, different choices of $\mathbf{D}$ will result in different $\mathbf{q}$ matrices. A sensible choice is a matrix whose $il$th component is

\begin{align}
D_{il}=\left\{\begin{array}{cc}
    \frac{1}{|\Omega_i|} & \text{if $S^{[l]}\in \Omega_i$},\\
    0 & \text{otherwise}, \end{array}\right.
\label{eq:22.50}
\end{align}

\noindent
which corresponds to a uniform averaging over microstates in a lumped state.
Using this matrix, called the \emph{distributor matrix}, in Eq. (\ref{eq:22.40}), can be shown to minimize $\| \mathbf{QC}-\mathbf{Cq}\|_{\rm F}$, the Frobenius norm of the \emph{approximate lumping discrepancy} $\mathbf{QC}-\mathbf{Cq}$ (see Ref. \cite{ward2022micro} for a proof), i.e., it is the closest to behaving like an exact lumping in this sense, for a given $\mathbf{C}$.

In what follows we will always use the approximate lumping generator matrix $\mathbf{q}$ determined by Eqs. (\ref{eq:22.40}) and (\ref{eq:22.50}):

\begin{align}
q_{ij} = (\mathbf{DQC})_{ij} = \sum_{k=1}^{M^N} D_{ik} (\mathbf{QC})_{kj} = \sum_{k=1}^{M^N} D_{ik} \sum_{S^{[l]} \in \Omega_j} Q_{kl} = \frac{1}{|\Omega_i|} \sum_{S^{[k]} \in \Omega_i} \sum_{S^{[l]} \in \Omega_j} Q_{kl}.
\label{eq:22.60}
\end{align}

\noindent
Thus $q_{ij}$ is the average of the sum of rates out of microstates in the $i$th lumped state and into the $j$th lumped state. The approximation enters only through Eq. (\ref{eq:22.60}): it arises from averaging the quantity $\rho_k = \sum_{S^{[l]} \in \Omega_j} Q_{kl}$ over the microstates $S^{[k]} \in \Omega_i$. Hence, the smaller the variation of $\rho_k$ within $\Omega_i$, the closer the lumping is to being exact.

\subsection{Generalized density-dependent population processes}
\label{sec23}

Consider a sequence of Markov chains indexed by a system-size parameter $N$. For each $N$, suppose that the state of the system is described by an $r$-dimensional vector whose components are non-negative integer counts. We assume that these counts are extensive, in the sense that they scale linearly with $N$.
[In the applications below, the components of the count vector are numbers of vertices and edges of different types in a graph; since the graphs considered are sparse, both vertex and edge counts are $\mathcal{O}(N)$.]
Because the counts are extensive, we may describe a state of the system by the corresponding density vector, obtained by dividing each count by $N$. Thus the set of possible density vectors, for a given $N$, may be written as


\begin{align}
\omega_N = \omega \cap \{ \mathbf{k} / N \, | \, \mathbf{k} \in \mathbb{Z}^r \},
\nonumber
\end{align}

\noindent
where $\omega \subset \mathbb{R}^r$ is the compact set of feasible density vectors. In the classical population-process setting, the components of the count vector are the numbers of individuals in each of $r$ possible states, and $\omega$ is the corresponding simplex in $\mathbb{R}^r$. Here we use the same density-dependent framework more generally: the components may be any integer-valued quantities that scale linearly with $N$, and $\omega$ is determined by the constraints that these quantities satisfy.
Let $\mathbf{q}^{(N)}$ be the infinitesimal generator for this Markov chain.
For a \emph{density-dependent process} there must exist a set of non-negative functions $\lambda_{\boldsymbol{\Delta}}: \omega \rightarrow \mathbb{R}_{\geq 0}$, one function for each possible \emph{jump vector} $\boldsymbol{\boldsymbol{\Delta}} \in \mathbb{Z}^r$, with the following property.
If the process is in state $\mathbf{x} \in \omega_N$, and a jump of type $\boldsymbol{\Delta}$ is possible (i.e., $\mathbf{x} + \boldsymbol{\Delta} / N \in \omega_N$), then the transition from $\mathbf{x}$ to $\mathbf{x} + \boldsymbol{\Delta} / N$ has rate

\begin{align}
q_{\mathbf{x},\mathbf{x} + \boldsymbol{\Delta} / N}^{(N)} = N \left[ \lambda_{\boldsymbol{\Delta}}(\mathbf{x}) + \mathcal{O} \left( \frac{1}{N} \right) \right],
\label{eq:23.20}
\end{align}



\noindent
so the $\lambda$ functions may be thought of as \emph{transition rate densities}. Importantly we require that the transition rate densities do not depend on $N$; any $N$-dependence in the transition rates $q_{\mathbf{x},\mathbf{x} + \boldsymbol{\Delta} / N}^{(N)}$ must be captured by the prefactor $N$ and the $\mathcal{O}(1/N)$ correction.
We say that the \emph{density-dependent family} corresponding to $\lambda_{\boldsymbol{\Delta}}$ is a sequence $\{\mathbf{Y}_N \, | \, N =1,2,3, \ldots \}$ of Markov processes such that the Markov process $\mathbf{Y}_N$ has state space $\omega_N$ and transition rates given by Eq. (\ref{eq:23.20}).
We define the \emph{drift field} $\mathbf{F}$ associated with the density-dependent family, as

\begin{align}
\mathbf{F}(\mathbf{y}) = \sum_{\boldsymbol{\Delta} \in \mathbb{Z}^r} \boldsymbol{\Delta} \lambda_{\boldsymbol{\Delta}}(\mathbf{y}),
\label{eq:23.30}
\end{align}

\noindent
for all $\mathbf{y} \in \omega$.
Intuitively, the quantity $\mathbf{F}(\mathbf{y})$ is the expected rate of change of the random variable $\mathbf{Y}_N(t)$, for large $N$, when $\mathbf{Y}_N(t) = \mathbf{y}$.
Let $\mathbf{y}(t)$ be the solution of

\begin{align}
\dot{\mathbf{y}} = \mathbf{F}(\mathbf{y}).
\label{eq:23.50}
\end{align}

\noindent
We assume that the initial conditions are chosen so that $\mathbf{Y}_N(0) \to \mathbf{y}(0)$ as $N \to \infty$.
Then, provided that for each compact $\psi \subset \omega$,

\begin{align}
\sum_{\boldsymbol{\Delta} \in \mathbb{Z}^r} |\boldsymbol{\Delta}| \sup_{y \in \psi} \lambda_{\boldsymbol{\Delta}} < \infty,
\label{eq:23.53}
\end{align}

\noindent
and that there exists a $K_\psi > 0$ such that

\begin{align}
|\mathbf{F}(\mathbf{x}) - \mathbf{F}(\mathbf{y})| \leq K_\psi |\mathbf{x}-\mathbf{y}|,
\label{eq:23.57}
\end{align}

\noindent
there is almost sure convergence between $\mathbf{Y}_N(t)$ and $\mathbf{y}(t)$ for any $t$ in the limit $N \to \infty$. (A more precise statement can be found in Ref. \cite{ethier2009markov}.)
This makes intuitive sense: as $N \to \infty$, the process makes increasingly many jumps of size $\sim 1/N$. As a result, fluctuations average out and $\mathbf{Y}_N(t)$ concentrates around a deterministic trajectory $\mathbf{y}(t)$, which is governed by Eq. (\ref{eq:23.50}).

Conditions (\ref{eq:23.53}) and (\ref{eq:23.57}) are standard regularity assumptions for density-dependent population processes. Condition (\ref{eq:23.53}) ensures that the total contribution of all possible jumps remains finite, preventing arbitrarily large or infinitely frequent transitions, while (\ref{eq:23.57}) is a local Lipschitz condition on the function $\mathbf{F}(\mathbf{y})$, ruling out excessively irregular dependence on the current state.
These conditions are mild and easily satisfied in the majority of dynamical processes studied on networks.
They are straightforward to verify in the models considered below.

Note that taking the limit $N \to \infty$ of density-dependent processes generally results in a significant dimensional reduction of the problem.
For finite $N$, the process is a Markov chain on the discrete state space $\omega_N$, and its full description requires the probability distribution over all feasible density vectors. The $N \to \infty$ limit replaces this stochastic description by the deterministic trajectory $\mathbf{y}(t)$, governed by the drift field $\mathbf{F}$. This is a substantial simplification: instead of solving the forward Kolmogorov equation for the full space of the finite-$N$ chain, one solves a much smaller system of ordinary differential equations for the density variables.




\section{Approximate lumping: Edge-based population models}
\label{sec3}

Our aim in this paper is to show how EBMF approximations, in the form of ordinary differential equations, for a wide range of dynamics on networks, can be systematically derived from the original, complete Markov chain description of the given model, by considering approximate lumping with appropriately chosen partitions and taking the large-$N$ limit. In particular we will consider approximate lumping that results in edge-based population models. In this case the lumping partition cells are sets of microstates that have the same number of vertices in each vertex state and the same number of edges in each \emph{edge state}. By edge state we mean an unordered pair $(\mathcal{A}, \mathcal{B})$ of vertex states corresponding to the vertex states of the two end vertices of the given edge. More precisely, for a given microstate $S \in \Omega$ the edge state of a given edge $(v_1,v_2)$ is $S(v_1,v_2) = (S(v_1), S(v_2))$.

\subsection{General case of $M$ vertex states}
\label{sec31}

For our general model, where vertices can be in $M$ different vertex states, there are $M + M(M-1)/2 = M(M+1)/2$ possible edge states, corresponding to the $M$ possible edge states where both vertices have the same state, and the $M(M-1)/2$ mixed edge states. The number of possible lumped states in the edge-based population model is the number of realisable configurations of numbers of vertices and edges in all the possible vertex states and edge states. This number strongly depends on the actual network structure. We can give a loose upper bound by assuming that all configurations of numbers that sum to $N$ for vertices and sum to $L$ for edges, are realisable. The number of ways $M$ vertex states can be distributed over $N$ vertices is

\begin{align}
\binom{N + M - 1}{M - 1},
\nonumber
\end{align}

\noindent
and the number of ways $M(M+1)/2$ edge states can be distributed over $L$ edges is

\begin{align}
\binom{L + M(M+1)/2 - 1}{M(M+1)/2 - 1},
\nonumber
\end{align}

\noindent
so we have, for the number of possible lumped states,

\begin{align}
n \leq \binom{N + M - 1}{M - 1} \binom{L + M(M+1)/2 - 1}{M(M+1)/2 - 1} = \mathcal{O} \left( N^{ \frac{ (M+4)(M-1) }{ 2 } } \right).
\label{eq:31.30}
\end{align}

\noindent
In writing Eq. (\ref{eq:31.30}) we assumed that our network is sparse, i.e., $L = \mathcal{O}(N)$, and we also assumed $M = \mathcal{O}(1)$. Thus edge-based lumping provides a reduction from the original $M^N$-dimensional Markov chain to an $N^{(M+4)(M-1)/2}$-dimensional Markov chain, which is a considerable difference when $N$ is large. In the $N \to \infty$ limit, assuming density dependence, the description of the lumped Markov chain reduces to a set of just $(M+4)(M-1)/2$ independent ODEs.

The details of our approach are somewhat involved, and presenting it for the full general dynamics, with $M$ vertex states, introduced in Sec. \ref{sec21}, would be unnecessarily lengthy and would likely obscure the main ideas. For this reason, to demonstrate our approach we will only consider binary-state dynamics, specifically, the case of the SISa model (Sec. \ref{sec32}). Through this example we will show how our approximate lumping method can be used to derive EBMF equations on a single graph (Sec. \ref{sec4}), and also on graph ensembles (Sec. \ref{sec5}). It is straightforward to extend the approach to more vertex states, however, writing the resulting equations in full generality would not be informative.

\subsection{Binary-state dynamics: SISa model}
\label{sec32}

As an example we consider SISa dynamics, where vertices can be in two states: \emph{susceptible} ($\textrm{S}$) and \emph{infected} ($\textrm{I}$). Consequently, edges can be in three possible states, depending on the states of the vertices they connect: \emph{susceptible-susceptible} ($\textrm{SS}$), \emph{susceptible-infected} ($\textrm{SI}$) and \emph{infected-infected} ($\textrm{II}$). A susceptible vertex with $d_{\textrm{I}}$ infected neighbours becomes infected at a rate $\alpha + d_{\textrm{I}} \beta$, and an infected vertex becomes susceptible at a rate $\gamma$.
This is a generalization of the standard SIS epidemiological model, incorporating a constant infection rate $\alpha$ that is independent of the neighbourhood. This ensures the existence of a non-trivial (endemic) steady state, since the all-susceptible configuration is no longer absorbing when $\alpha > 0$.
We define our approximate lumping partition $\Pi_\Omega = \{ \Omega_1, \Omega_2, \ldots \Omega_n \}$ so that the partition cells are sets of microstates that have the same number of vertices in each vertex state and the same number of edges in each edge state. One such lumping partition cell, and the corresponding lumped state, may be described by $\mathbf{s}\in\mathbb{Z}^{5}_{\ge0}$, whose components are the numbers of vertices and edges in the possible vertex and edge states, $N_{\textrm{I}}, N_{\textrm{S}}, N_{\textrm{II}}, N_{\textrm{SI}}, N_{\textrm{SS}}$, respectively.
Note that, since $N_{\textrm{I}}+N_{\textrm{S}}=N$ and $N_{\textrm{II}}+N_{\textrm{SI}}+N_{\textrm{SS}}=L$, three of the five parameters---any one of the vertex-state counts and any two of the edge-state counts---are sufficient to describe a lumped state; however, retaining the full notation makes the derivation clearer.

\begin{figure}[H]
\centering
\includegraphics[width=1.0\textwidth]{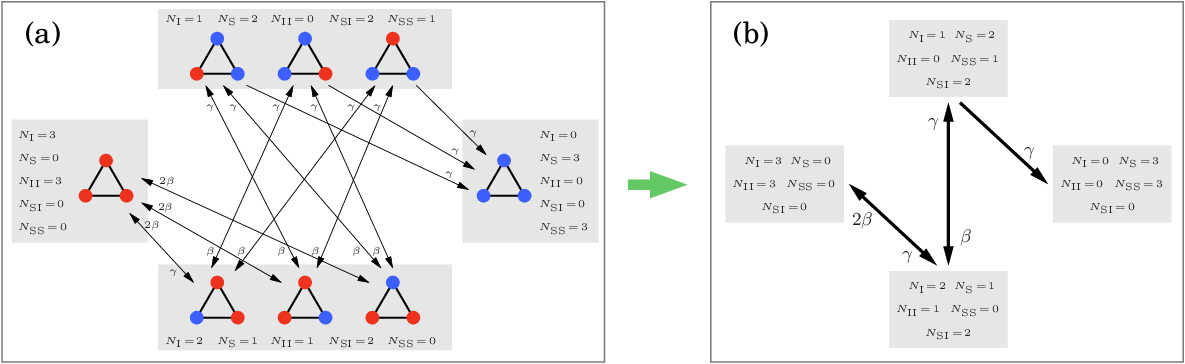}
\caption{Edge-based lumping of the SIS model (equivalent to the SISa model with $\alpha=0$) on a triangle graph. (a) The $2^N = 8$ microstates (red=infected, blue=susceptible) are partitioned into four edge-based lumping partitions (grey boxes). Arrows indicate the possible transitions between microstates, with the transition rates shown next to the arrows [elements of matrix $\mathbf{Q}$, according to Eq. (\ref{eq:32.20})]. According to the criterion, see Sec. \ref{sec22}, edge-based lumping in this case constitutes an exact lumping. (b) Transitions and transition rates [elements of matrix $\mathbf{q}$, according to Eq. (\ref{eq:32.30})] between lumped states (grey boxes), in the edge-based lumping of SIS dynamics on a triangle graph.}
\label{fig:EB_lumping}
\end{figure}

We enumerate the $n$ lumped states in some way, and denote them by $\mathbf{s}^{[1]}, \mathbf{s}^{[2]}, \ldots, \mathbf{s}^{[n]}$.
According to SISa dynamics, a given microstate in $\mathbf{s}^{[i]}$ may transition into multiple other lumped states, depending on the number and vertex states of the neighbours of the transition vertex [see Fig. \ref{fig:EB_lumping}(a) for an example of lumping SIS dynamics on the triangle graph]. Specifically, if the transition vertex $v$ has $d_{\textrm{I}}$ infected and $d_{\textrm{S}}$ susceptible neighbours, then the corresponding transition, from a microstate in $\mathbf{s}^{[i]}$, will be to a microstate in one of two lumped states: $\mathbf{s}^{[i]} + \boldsymbol{\Delta}_{d_{\textrm{I}}, d_{\textrm{S}}}^+$ or $\mathbf{s}^{[i]} + \boldsymbol{\Delta}_{d_{\textrm{I}}, d_{\textrm{S}}}^-$, where

\begin{align}
\boldsymbol{\Delta}_{d_{\textrm{I}}, d_{\textrm{S}}}^+ &= (1, -1, d_{\textrm{I}}, d_{\textrm{S}}-d_{\textrm{I}}, -d_{\textrm{S}})
\label{eq:32.10a}
\end{align}

\noindent
if vertex $v$ made the vertex state transition $\textrm{S} \to \textrm{I}$, and

\begin{align}
\boldsymbol{\Delta}_{d_{\textrm{I}}, d_{\textrm{S}}}^- &= (-1, 1, -d_{\textrm{I}}, d_{\textrm{I}}-d_{\textrm{S}}, d_{\textrm{S}})
\label{eq:32.10b}
\end{align}

\noindent
if vertex $v$ made the vertex state transition $\textrm{I} \to \textrm{S}$. The jump vectors $\boldsymbol{\Delta}_{d_{\textrm{I}}, d_{\textrm{S}}}^+$ and $\boldsymbol{\Delta}_{d_{\textrm{I}}, d_{\textrm{S}}}^-$ describe the only possible transitions to a new lumped state when the transition vertex has $d_{\textrm{I}}$ infected and $d_{\textrm{S}}$ susceptible neighbours.

Let us consider a transition from microstate $S^{[k]} \in \Omega_i$ into microstate $S^{[l]} \in \Omega_j$. 
From the microstates we can identify the corresponding pair of lumped states $(\mathbf{s}^{[i]}, \mathbf{s}^{[j]})$, but from this pair alone we can also determine the numbers $d_{\textrm{I}}, d_{\textrm{S}}$ and also the vertex state $\mathcal{A}$ (either $\textrm{S}$ or $\textrm{I}$) of the transition vertex in microstate $S^{[k]}$.
Thus, for a given pair $(\mathbf{s}^{[i]}, \mathbf{s}^{[j]})$ the transition vertex can be any vertex $v$ for which there exists at least one microstate in $\Omega_i$ where vertex $v$ has vertex state $\mathcal{A}$, has $d_{\textrm{I}}$ infected neighbours and $d_{\textrm{S}}$ susceptible neighbours. To calculate $q_{ij}$ we must sum all the transition rates corresponding to different pairs $(S^{[k]}, S^{[l]})$ where $S^{[k]} \in \Omega_i$ and $S^{[l]} \in \Omega_j$. Note that the transition rates between all transition pairs corresponding to the lumped states $(\mathbf{s}^{[i]}, \mathbf{s}^{[j]})$ are identical, in other words, $Q_{kl}$ only depends on $i,j$:
\begin{align}
Q_{kl} =
\begin{cases}
\alpha + d_{\textrm{I}}(i,j) \beta \quad & \mathcal{A}(i,j) = \textrm{S}, \\
\gamma \quad & \mathcal{A}(i,j) = \textrm{I},
\end{cases}
\label{eq:32.20}
\end{align}

\noindent
where we indicated that the number of infected neighbours $d_{\textrm{I}}$, and the original vertex state $\mathcal{A}$ of the transition vertex $v$ are functions of $i,j$.
Thus edge-based lumping is exact when all microstates in each $\Omega_{i}$ have the same number of transition pairs to any other $\Omega_{j}$; however, this is not typically the case.

Whether lumping is exact or not, the double sum over microstates in Eq. (\ref{eq:22.60}) may be rewritten as a single sum over vertices,




\begin{align}
q_{ij} =
\begin{dcases}
\frac{1}{\mathcal{N}(\mathbf{s}^{[i]})} \sum_{v \in V} \mathcal{N}^v \big( \textrm{S}, \mathbf{d}(i,j), \mathbf{s}^{[i]} \big)\big( \alpha + d_{\textrm{I}}(i,j) \, \beta \big) \quad & \mathcal{A}(i,j) = \textrm{S}, \\
\frac{1}{\mathcal{N}(\mathbf{s}^{[i]})} \sum_{v \in V} \mathcal{N}^v \big( \textrm{I}, \mathbf{d}(i,j), \mathbf{s}^{[i]} \big) \gamma \quad & \mathcal{A}(i,j) = \textrm{I},
\end{dcases}
\label{eq:32.30}
\end{align}

\noindent
where $\mathcal{N}(\mathbf{s}^{[i]})=|\Omega_i|$, $\mathbf{d}(i,j) = ( d_{\textrm{I}}(i,j), d_{\textrm{S}}(i,j) )$, and
$\mathcal{N}^v (\mathcal{A}, \mathbf{d}, \mathbf{s})$ 
is the number of microstates corresponding to the lumped state $\mathbf{s}$ in which vertex $v$ is in vertex state $\mathcal{A}$ and has neighbourhood $\mathbf{d}=(d_{\textrm{I}},d_{\textrm{S}})$.
Eq. (\ref{eq:32.30}) follows by reorganizing the double sum in Eq. (\ref{eq:22.60}) according to the choice of transition vertex.
For fixed lumped states $\mathbf{s}^{[i]}$ and $\mathbf{s}^{[j]}$, a transition from a microstate in $\Omega_i$ to a microstate in $\Omega_j$ can only occur when some vertex $v$ in a microstate in $\Omega_i$ has the unique local configuration specified by $\mathcal{A}(i,j)$, $d_{\textrm{I}}(i,j)$, and $d_{\textrm{S}}(i,j)$.
Since all such transitions have the same rate in a homogeneous SVT model, the total contribution is obtained by multiplying that rate by the number of microstates in $\Omega_i$ for which vertex $v$ has that configuration, then summing over all vertices $v$. Dividing by $|\Omega_i|$ gives the uniform average over microstates in the lumped state, consistent with the approximate lumping defined in Eq. (\ref{eq:22.60}).
[See Fig. \ref{fig:EB_lumping}(b) for an example of the lumped system---using Eq. (\ref{eq:32.30})---for SIS dynamics on the triangle graph.]
Naturally, as with any infinitesimal generator, we must have

\begin{align}
q_{ii} = - \sum_{j \neq i} q_{ij}
\nonumber
\end{align}

\noindent
for all $i$.

The number of microstates in binary-state dynamics is $2^N$, therefore a full description of the SISa model on a network of $N$ vertices would require the solution of $2^N$ equations. The number of lumped states is now, according to Eq. (\ref{eq:31.30}),

\begin{align}
n = \mathcal{O}(N^3),
\nonumber
\end{align}

\noindent
assuming, as before, that the network is sparse.
Thus we only need to solve $\sim N^3$ equations to fully describe the edge-based population model based on an approximate lumping, which is a significant dimensional reduction.
The number of equations required to describe the system is drastically reduced even further in the large-$N$ limit, if the model is density-dependent (see Sec. \ref{sec23}), i.e., the transition rates, Eq. (\ref{eq:32.30}), between lumped states can be written in the form


\begin{align}
q_{\mathbf{s}, \mathbf{s} + \boldsymbol{\Delta}} = N \left[ \lambda_{\boldsymbol{\Delta}} \left( \frac{\mathbf{s}}{N} \right) + \mathcal{O} \left( \frac{1}{N}  \right)  \right].
\label{eq:32.40}
\end{align}

\noindent
In this case, in the large-$N$ limit, the evolution of the dynamics is deterministic, and given by the system of differential equations

\begin{align}
\dot{\mathbf{y}} = \mathbf{F} ( \mathbf{y} ),
\label{eq:32.50}
\end{align}

\noindent
where $\mathbf{y} = (y_{\textrm{I}}, y_{\textrm{S}}, y_{\textrm{II}}, y_{\textrm{SI}}, y_{\textrm{SS}})$ is the density of vertices and edges in the given states and

\begin{align}
\mathbf{F} ( \mathbf{y} ) = \sum_{\boldsymbol{\Delta}} \boldsymbol{\Delta} \lambda_{\boldsymbol{\Delta}}(\mathbf{y}) = \sum_{d_{\textrm{I}}=0}^{k_{\textrm{max}}} \sum_{d_{\textrm{S}}=0}^{k_{\textrm{max}}} \left[ \boldsymbol{\Delta}_{d_{\textrm{I}},d_{\textrm{S}}}^+ \lambda_{\boldsymbol{\Delta}_{d_{\textrm{I}},d_{\textrm{S}}}^+}(\mathbf{y}) + \boldsymbol{\Delta}_{d_{\textrm{I}},d_{\textrm{S}}}^- \lambda_{\boldsymbol{\Delta}_{d_{\textrm{I}},d_{\textrm{S}}}^-}(\mathbf{y}) \right],
\label{eq:32.60}
\end{align}

\noindent
as described in Sec. \ref{sec23}. The upper limit $k_{\textrm{max}}$ in the sums in Eq. (\ref{eq:32.60}) is the maximum degree.

The main challenge in applying our method to a given network lies in the combinatorial evaluation of $\mathcal{N}(\mathbf{s})$ and $\mathcal{N}^v (\mathcal{A}, \mathbf{d}, \mathbf{s})$, which determine the transition rate densities, see Eq. (\ref{eq:32.30}).
Aside from this step, the procedure---defining the lumping, constructing the transition rates via Eq. (\ref{eq:32.30}), and taking the large-$N$ limit to obtain the mean-field ODEs---is systematic and applies in the same way to both single graphs and graph ensembles. 
We will demonstrate this in Sec. \ref{sec4} for the case of a cycle, and in Sec. \ref{sec5} for the case of random regular networks and Erd\H{o}s-R\'enyi networks.

\subsection{Hypergeometric distribution}
\label{sec33}

We will make extensive use of the hypergeometric distribution, which is defined as

\begin{align}
H(k; N, K, n) = \frac{ \binom{K}{k} \binom{N-K}{n-k} }{ \binom{N}{n} },
\label{eq:33.10}
\end{align}

\noindent
with $0 \leq k \leq \min(n,K)$. The hypergeometric distribution describes the probability of obtaining $k$ successes in $n$ trials from a population of $N$, of which $K$ members could be successes. One can use Vandermonde's identity to verify that $H$ is correctly normalized:

\begin{align}
\sum_{k=0}^{\min(n,K)} \binom{K}{k} \binom{N-K}{n-k} = \binom{N}{n}.
\nonumber
\end{align}

\noindent
In the derivations below we will often refer to the first two moments of this distribution, which we write here for reference:

\begin{align}
\sum_{k=0}^{\min(n,K)} k H(k; N, K, n) &= n \frac{K}{N},  \label{eq:33.31}  \\
\sum_{k=0}^{\min(n,K)} k^2 H(k; N, K, n) &= n \frac{K}{N} + n(n-1) \frac{ K(K-1) }{ N(N-1) }.   \label{eq:33.32}
\end{align}

\section{Application to a single graph: cycle}
\label{sec4}

In this section we will use the example of the cycle graph to demonstrate how our approach may be applied to single graphs.

For degree-regular graphs, where each vertex has degree $c$, we have that 
$c N_{\textrm{I}} = N_{\textrm{SI}} + 2 N_{\textrm{II}}$ , i.e. the number of stubs emanating from infected vertices is equal to the number of susceptible-infected edges plus twice the number of infected-infected edges. (Similar reasoning yields $c N_{\textrm{S}} = N_{\textrm{SI}} + 2 N_{\textrm{SS}}$.)
Thus, for degree-regular graphs only two parameters are sufficient to describe a lumped state, e.g. $N_{\textrm{I}}$ and $N_{\textrm{SI}}$, though for consistency of notation we will continue to use $\mathbf{s}$ to refer to the lumped state. Furthermore, for degree-regular graphs we have $c=d_{\textrm{I}}+d_{\textrm{S}}$.

In the case of a cycle, $c=2$, one can obtain explicit formulas for the microstate counts required. The number $\mathcal{N}(\mathbf{s})$ of microstates in the lumped state characterized by $\mathbf{s}$ is

\begin{align}
\mathcal{N}(\mathbf{s}) =
\begin{cases}
\frac{N N_{\textrm{SI}}}{2 N_{\textrm{I}} N_{\textrm{S}}} \binom{N_{\textrm{I}}}{N_{\textrm{SI}}/2} \binom{N_{\textrm{S}}}{N_{\textrm{SI}}/2} \quad & 1 \leq N_{\textrm{I}} \leq N-1 \\
1 \quad & N_{\textrm{I}}=0 \quad \text{or} \quad N_{\textrm{I}}=N,
\end{cases}
\label{eq:4.10}
\end{align}

\noindent
for all realisable values of $N_{\textrm{SI}}$. The derivation of Eq. (\ref{eq:4.10}) is presented in Appendix \ref{secA1}.
The numbers of microstates corresponding to $\mathbf{s}$ in which the transition vertex $v$ has neighbourhood $\mathbf{d}=(d_{\textrm{I}},d_{\textrm{S}})$ and $v$ is susceptible are
\begin{align}
\mathcal{N}^v(\textrm{S}, \mathbf{d}, \mathbf{s}) =
\begin{cases}
\binom{N_{\textrm{I}}-1}{ N_{\textrm{SI}}/2 -1} \binom{2}{d_{\textrm{I}}} \binom{N_{\textrm{S}}-2}{N_{\textrm{SI}}/2-d_{\textrm{I}}} \quad & 1 \leq N_{\textrm{I}} \leq N-2 \\
1 \quad & N_{\textrm{I}} = 0 \quad \text{or} \quad N_{\textrm{I}}=N-1,
\end{cases}
\label{eq:4.15}
\end{align}

\noindent
and similarly when $v$ is infected we have

\begin{align}
\mathcal{N}^v(\textrm{I}, \mathbf{d}, \mathbf{s}) =
\begin{cases}
\binom{N_{\textrm{S}}-1}{ N_{\textrm{SI}}/2 -1} \binom{2}{2-d_{\textrm{I}}} \binom{N_{\textrm{I}}-2}{N_{\textrm{SI}}/2-(2-d_{\textrm{I}})} \quad & 2 \leq N_{\textrm{I}} \leq N-1 \\
1 \quad & N_{\textrm{I}} = 1 \quad \text{or} \quad N_{\textrm{I}}=N,
\end{cases}
\label{eq:4.17}
\end{align}

\noindent
for all realisable values of $d_{\textrm{I}}$ and $N_{\textrm{SI}}$. The derivations of Eqs. (\ref{eq:4.15}) and (\ref{eq:4.17}) are presented in Appendix \ref{secA2}.

The possible transitions between different lumped states are characterized by the jump vectors $\boldsymbol{\Delta}_{d_{\textrm{I}}, d_{\textrm{S}}}^+$ and $\boldsymbol{\Delta}_{d_{\textrm{I}}, d_{\textrm{S}}}^-$. In the case of degree-regular graphs, and hence also cycles, where $d_{\textrm{I}}$ and $d_{\textrm{S}}$ are not independent, we can simplify this notation and simply use $\boldsymbol{\Delta}_{d_{\textrm{I}}}^+$ and $\boldsymbol{\Delta}_{d_{\textrm{I}}}^-$, which, in the case of cycles, are

\begin{align}
\boldsymbol{\Delta}_{d_{\textrm{I}}}^+ &= (1, -1, d_{\textrm{I}}, 2-2d_{\textrm{I}}, d_{\textrm{I}}-2),  \nonumber \\ 
\boldsymbol{\Delta}_{d_{\textrm{I}}}^- &= (-1, 1, -d_{\textrm{I}}, 2d_{\textrm{I}}-2, 2-d_{\textrm{I}}).  \nonumber 
\end{align}

\noindent
When lumped state $\mathbf{s}$ is such that $N_{\textrm{I}} \leq N-2$, we can write the transition rate from $\mathbf{s}$ into lumped state $\mathbf{s}+\boldsymbol{\Delta}_{d_{\textrm{I}}}^+$, corresponding to an infection, using Eq. (\ref{eq:32.30}),

\begin{align}
q_{\mathbf{s},\mathbf{s}+\boldsymbol{\Delta}_{d_{\textrm{I}}}^+} &= \frac{ \alpha + d_{\textrm{I}} \beta }{ \mathcal{N}(\mathbf{s}) } \sum_{v \in V} \mathcal{N}^v(\textrm{S}, \mathbf{d}, \mathbf{s})   \nonumber \\
&= (\alpha + d_{\textrm{I}} \beta) \, N \frac{ \mathcal{N}^v(\textrm{S}, \mathbf{d}, \mathbf{s}) }{ \mathcal{N}(\mathbf{s}) }   \nonumber \\
&= (\alpha + d_{\textrm{I}} \beta) \, N_{\textrm{S}} \frac{ \binom{2}{d_{\textrm{I}}} \binom{N_{\textrm{S}}-2}{N_{\textrm{SI}}/2-d_{\textrm{I}}} }{ \binom{N_{\textrm{S}}}{N_{\textrm{SI}}/2} }   \nonumber \\
&= (\alpha + d_{\textrm{I}} \beta) \, N_{\textrm{S}} H(d_{\textrm{I}}; N_{\textrm{S}}, 2, N_{\textrm{SI}}/2),
\label{eq:4.20}
\end{align}

\noindent
for all realisable values of $d_{\textrm{I}}$ and $N_{\textrm{SI}}$. To obtain Eq. (\ref{eq:4.20}) we used the symmetry with respect to choosing the transition vertex $v$, and used Eqs. (\ref{eq:4.15}), (\ref{eq:4.10}) and (\ref{eq:33.10}). For the boundary case of $N_{\textrm{I}} = N-1$ we have $q_{\mathbf{s},\mathbf{s}+\boldsymbol{\Delta}_{d_{\textrm{I}}}^+} = \alpha + 2 \beta$ and for $N_{\textrm{I}}=N$ clearly $q_{\mathbf{s},\mathbf{s}+\boldsymbol{\Delta}_{d_{\textrm{I}}}^+} = 0$ as there are no more vertices to become infected.
In completely analogous fashion, when $N_{\textrm{I}} \geq 2$, we can write the transition rate from lumped state $\mathbf{s}$ into lumped state $\mathbf{s}+\boldsymbol{\Delta}_{d_{\textrm{I}}}^-$, corresponding to a recovery,

\begin{align}
q_{\mathbf{s},\mathbf{s}+\boldsymbol{\Delta}_{d_{\textrm{I}}}^-} &= \frac{ \gamma }{ \mathcal{N}(\mathbf{s}) } \sum_{v \in V} \mathcal{N}^v(\textrm{I}, \mathbf{d}, \mathbf{s})   \nonumber \\
&= \gamma N \frac{ \mathcal{N}^v(\textrm{I}, \mathbf{d}, \mathbf{s}) }{ \mathcal{N}(\mathbf{s}) }   \nonumber \\
&= \gamma N_{\textrm{I}} \frac{ \binom{2}{2-d_{\textrm{I}}} \binom{N_{\textrm{I}}-2}{N_{\textrm{SI}}/2-(2-d_{\textrm{I}})} }{ \binom{N_{\textrm{I}}}{N_{\textrm{SI}}/2} }   \nonumber \\
&= \gamma N_{\textrm{I}} H(2-d_{\textrm{I}}; N_{\textrm{I}}, 2, N_{\textrm{SI}}/2),
\label{eq:4.30}
\end{align}

\noindent
for all realisable values of $d_{\textrm{I}}$ and $N_{\textrm{SI}}$. For the boundary case of $N_{\textrm{I}} = 1$ we have $q_{\mathbf{s},\mathbf{s}+\boldsymbol{\Delta}_{d_{\textrm{I}}}^-} = \gamma$ and for $N_{\textrm{I}}=0$ clearly $q_{\mathbf{s},\mathbf{s}+\boldsymbol{\Delta}_{d_{\textrm{I}}}^-} = 0$ as there are no more vertices to recover.
The lumped transition rates, Eqs. (\ref{eq:4.20}) and (\ref{eq:4.30}), represent an approximation to the original dynamics.
One can verify that these rates are density-dependent, i.e., have the form of Eq. (\ref{eq:32.40}).
To obtain the ODEs describing the $N \to \infty$ limit of this approximate dynamics, we must write the function $\mathbf{F}(\mathbf{y})$, see Eq. (\ref{eq:32.60}),

\begin{align}
\mathbf{F} ( \mathbf{y} ) &= \sum_{d_{\textrm{I}}=0}^2 \left[ \boldsymbol{\Delta}_{d_{\textrm{I}}}^+ \frac{   q_{\mathbf{s},\mathbf{s}+\boldsymbol{\Delta}_{d_{\textrm{I}}}^+}   }{ N } + \boldsymbol{\Delta}_{d_{\textrm{I}}}^- \frac{  q_{\mathbf{s},\mathbf{s}+\boldsymbol{\Delta}_{d_{\textrm{I}}}^-}  }{ N } \right]  \nonumber \\
&= \beta y_{\textrm{S}} \sum_{d_{\textrm{I}}=0}^2 \boldsymbol{\Delta}_{d_{\textrm{I}}}^+ d_{\textrm{I}} H(d_{\textrm{I}}; N_{\textrm{S}}, 2, N_{\textrm{SI}}/2) + \alpha y_{\textrm{S}} \sum_{d_{\textrm{I}}=0}^2 \boldsymbol{\Delta}_{d_{\textrm{I}}}^+ H(d_{\textrm{I}}; N_{\textrm{S}}, 2, N_{\textrm{SI}}/2)  \nonumber \\
& \quad + \gamma y_{\textrm{I}} \sum_{d_{\textrm{I}}=0}^2 \boldsymbol{\Delta}_{d_{\textrm{I}}}^- H(2-d_{\textrm{I}}; N_{\textrm{I}}, 2, N_{\textrm{SI}}/2),
\label{eq:4.40}
\end{align}

\noindent
where $\mathbf{y} = (y_{\textrm{I}}, y_{\textrm{S}}, y_{\textrm{II}}, y_{\textrm{SI}}, y_{\textrm{SS}})$. This normalization is natural in the case of a cycle, since the number of vertices is equal to the number of edges, i.e., $y_{\textrm{I}} + y_{\textrm{S}} = y_{\textrm{II}} + y_{\textrm{SI}} + y_{\textrm{SS}} = 1$.
In degree-regular networks $d_{\textrm{I}}$ and $d_{\textrm{S}}$ are not independent, therefore a single sum---e.g., over $d_{\textrm{I}}$---is required to write $\mathbf{F} ( \mathbf{y} )$, instead of the double sum in the more general form, Eq. (\ref{eq:32.60}).
Note also that in writing Eq. (\ref{eq:4.40}) we only considered the ``interior'' of the lumped state space, represented by Eqs. (\ref{eq:4.20}) and (\ref{eq:4.30}), disregarding the pathological boundary values $y_{\textrm{I}}=0$ and $y_{\textrm{I}}=1$.
The sums in Eq. (\ref{eq:4.40}), for each component of the jump vectors, are simply averages over the hypergeometric distribution, involving only the first and second moments. Using Eqs. (\ref{eq:33.31}) and (\ref{eq:33.32}) we can write these averages directly.
The resulting differential equations, using Eq. (\ref{eq:32.50}), for $0 < y_{\textrm{I}} < 1$, are

\begin{align}
\dot{y}_{\textrm{I}} &= \alpha y_{\textrm{S}} + \beta \left( \frac{ y_{\textrm{SI}} y_{\textrm{SS}} }{ y_{\textrm{S}} } + \frac{ y_{\textrm{SI}}^2 }{ 2 y_{\textrm{S}} } \right) - \gamma \left( \frac{ y_{\textrm{SI}}^2 }{ 4 y_{\textrm{I}} } + \frac{ y_{\textrm{SI}} y_{\textrm{II}} }{ y_{\textrm{I}} } + \frac{ y_{\textrm{II}}^2 }{ y_{\textrm{I}} } \right) \label{eq:4.111} \\
\dot{y}_{\textrm{S}} &= -\alpha y_{\textrm{S}} -\beta \left( \frac{ y_{\textrm{SI}} y_{\textrm{SS}} }{ y_{\textrm{S}} } + \frac{ y_{\textrm{SI}}^2 }{ 2 y_{\textrm{S}} } \right) + \gamma \left( \frac{ y_{\textrm{SI}}^2 }{ 4 y_{\textrm{I}} } + \frac{ y_{\textrm{SI}} y_{\textrm{II}} }{ y_{\textrm{I}} } + \frac{ y_{\textrm{II}}^2 }{ y_{\textrm{I}} } \right) \label{eq:4.112} \\
\dot{y}_{\textrm{II}} &= \alpha y_{\textrm{SI}} + \beta \left( \frac{ y_{\textrm{SI}} y_{\textrm{SS}} }{ y_{\textrm{S}} } + \frac{ y_{\textrm{SI}}^2 }{ y_{\textrm{S}} } \right) - \gamma \left( \frac{ y_{\textrm{SI}} y_{\textrm{II}} }{ y_{\textrm{I}} } + \frac{ 2 y_{\textrm{II}}^2 }{ y_{\textrm{I}} } \right) \label{eq:4.113} \\
\dot{y}_{\textrm{SI}} &= \alpha ( 2 y_{\textrm{SS}} - y_{\textrm{SI}} ) -\beta \frac{ y_{\textrm{SI}}^2 }{ y_{\textrm{S}} } - \gamma \left( \frac{ y_{\textrm{SI}}^2 }{ 2 y_{\textrm{I}} } - \frac{ 2 y_{\textrm{II}}^2 }{ y_{\textrm{I}} } \right) \label{eq:4.114} \\
\dot{y}_{\textrm{SS}} &= -2 \alpha y_{\textrm{SS}} - \beta \frac{ y_{\textrm{SI}} y_{\textrm{SS}} }{ y_{\textrm{S}} } + \gamma \left( \frac{ y_{\textrm{SI}}^2 }{ 2 y_{\textrm{I}} } + \frac{ y_{\textrm{SI}} y_{\textrm{II}} }{ y_{\textrm{I}} } \right). \label{eq:4.115}
\end{align}

\noindent
Using $N_{\textrm{SI}} = 2 N_{\textrm{I}} - 2 N_{\textrm{II}}$, and due to symmetry, also $N_{\textrm{SI}} = 2 N_{\textrm{S}} - 2 N_{\textrm{SS}}$, which are true for cycles, one can show that these differential equations are equivalent to the homogeneous pairwise mean-field approximation in Eqs. (4.10a--4.10e) of Ref. \cite{kiss2017book} derived, using moment closure arguments, for regular networks of degree $n$ in the notation of Ref. \cite{kiss2017book}, when substituting $n=2$.

It is useful to pause here and compare our derivation with the usual moment-closure route to pairwise mean-field equations. In a moment-based derivation, one first writes exact equations for lower-order quantities, such as vertex and edge counts, and then encounters higher-order terms describing the local arrangement of states around vertices. A closure assumption is then imposed to express these higher-order quantities in terms of the retained lower-order ones. In the present approach no such hierarchy is introduced, and no separate closure formula is postulated. Nevertheless, the same loss of information occurs: by lumping microstates according only to vertex and edge counts, we discard the detailed arrangement of states around vertices. The approximate lumping step replaces the exact dependence on this discarded information by an average over all microstates compatible with the retained counts. Thus we recover a well-known EBMF approximation using a systematic, principled approach that explicitly identifies the coarse-graining and averaging step where the approximation enters.

\vspace{1em}

\textbf{Accuracy of lumped Markov chain description vs mean-field ODEs.}--- It is instructive to compare the accuracy of the $\mathbf{q}$-matrix-description [Eqs. (\ref{eq:4.20}--\ref{eq:4.30})] of SISa dynamics on a finite-size cycle with that of the resulting mean-field description via ODEs [Eqs. (\ref{eq:4.111}--\ref{eq:4.115})]. We consider the steady state value $y_{\textrm{I}}^{\text{(st.)}}$ of the relative number of infected vertices. In the finite lumped system this value is easily expressed as a function of the stationary distribution of the lumped Markov chain, i.e., the solution of the equation $\mathbf{q}^T \mathbf{p} = 0$. In the EBMF system we can find the steady state value by setting the derivatives to zero in Eqs. (\ref{eq:4.111}--\ref{eq:4.115}) and solving. In this case we obtain the explicit formula


\begin{align}
y_{\textrm{I}}^{(\text{st., SISa, EBMF})} = 1 - \frac{\gamma}{ \beta + \sqrt{ (\alpha+\beta+\gamma)^2 - 4 \beta\gamma } }.
\label{eq:4.120}
\end{align}

\begin{figure}[H]
\centering
\includegraphics[width=0.9\textwidth]{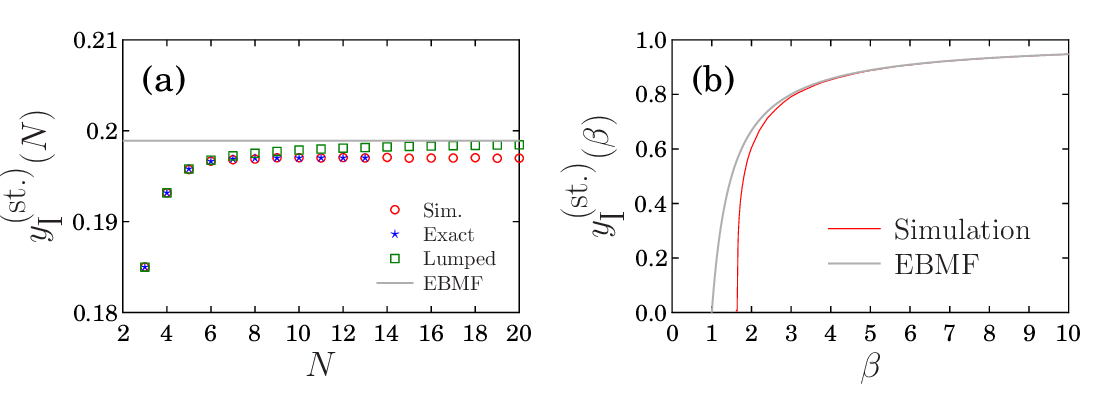}
\caption{(a) Simulation results for the steady state of the relative number of infected vertices as a function of system size $N$, for a range of small values of $N$, in the SISa model on a cycle, with parameters $\alpha = 0.1, \beta = 0.5, \gamma = 1$. The solutions of the full Markov chain, the lumped Markov chain, and the EBMF ODE system are shown for comparison. (b) Simulation results for the steady state value of the relative number of infected vertices as a function of the infection rate parameter $\beta$, for the SIS model on a cycle of $N=10000$. (Note that for the SIS model, for any finite value of $N$, this is only a quasi-stationary state.) The EBMF result is shown for comparison.}
\label{fig:steady_state}
\end{figure}

\noindent
In Fig. \ref{fig:steady_state}(a) we present the steady state value $y_{\textrm{I}}^{\text{(st.)}}$ in simulations of SISa dynamics on cycles of sizes up to $N=20$. The value of $y_{\textrm{I}}^{\text{(st.)}}$, acquired from the solution of $\mathbf{Q}^T \mathbf{P}=0$, using the infinitesimal generator of the full Markov chain, is also shown for sizes up to $N=13$ and it coincides perfectly with the simulation results. Solutions using the lumped generator coincide with the exact result for sizes $N=3,4,5$, as the lumping is exact in these cases. The lumping is exact for these sizes
because in these cases all microstates within a lumping partition are isomorphic and thus have identical transition rates into any other lumped state. This is not true for $N > 5$ and the lumped solution visibly diverges from the simulation results for larger cycle sizes. The lumped solution approaches the EBMF value [grey line in Fig. \ref{fig:steady_state}(a)], Eq. (\ref{eq:4.120}), for large-$N$.

\noindent
The lumped system, and even the EBMF, give good approximations for the SISa model. Mean-field descriptions may work very well away from criticality, and this appears to be true even in the case of an extreme low-dimensional object, a cycle. Near a critical point, however, mean-field approximations are expected to be inaccurate for low-dimensional systems. This is clearly seen in the SIS model, a special case of the SISa model, with $\alpha=0$, where the steady-state solution for the relative number of infected vertices simplifies to

\begin{align}
y_{\textrm{I}}^{(\text{st., SIS, EBMF})} = \frac{ 2 (\beta-\gamma) }{ 2 \beta - \gamma },
\nonumber
\end{align}

\noindent
which predicts a phase transition at the critical infection rate

\begin{align}
\beta_c^{(\text{SIS, EBMF})} = \gamma.
\nonumber
\end{align}

\noindent
Figure \ref{fig:steady_state}(b) shows that the EBMF approximation significantly underestimates the actual critical threshold, $\beta_c^{(\text{SIS})} \approx 1.649 \gamma$ (see, e.g., \cite{enss2004ageing}). Away from criticality, for higher values of $\beta$, the EBMF approximation is again highly accurate.

\section{Application to graph ensembles}
\label{sec5}

Our approach is straightforward to formulate for single networks, but performing the necessary microstate counts can become challenging unless the network has sufficient symmetry.
Another path forward to make use of our method is to apply it to graph ensembles which, with the appropriate structural properties, may still provide good approximations to dynamics on real networks.
To do so, we define the underlying Markov chain on a generalised microstate space, where each microstate specifies both the vertex states and a particular graph realization from the ensemble. The graph structure is fixed during the dynamics, and lumping is performed over this generalised space using the same edge-based variables as before. This allows us to compute the lumped transition rates by averaging over all graph realizations consistent with a given lumped state.
In this section we demonstrate how this can be done in the case of random regular networks and Erd\H{o}s-R\'enyi random networks.

\subsection{Random regular graph ensemble}
\label{sec51}

Let us apply our approach to the case of random regular networks. More specifically, we consider $N$ labelled vertices, each with degree $c$. Once again, each vertex can be in one of two states, $\textrm{S}$ or $\textrm{I}$ and each edge in one of three states, $\textrm{SS}$, $\textrm{II}$ or $\textrm{SI}$. We will consider the configuration model ensemble, where we assume that each vertex has a prescribed number (its degree; in this case $c$ for all vertices) of stubs emanating from it, and all possible stub wiring configurations must be considered with equal weight. Thus we consider the stubs of each vertex to be labelled, in addition to the vertices themselves being labelled.
Let us define a Markov chain whose microstates are specified by $(\boldsymbol{\sigma}, \mathcal{C})$, where $\boldsymbol{\sigma}$ is an $N$-tuple indicating the vertex state of each individual vertex and $\mathcal{C}$ is a configuration of stub wirings. There will be many stub wirings that correspond to the same labelled graph (when considering only vertex labels).
In fact, each distinct simple labelled graph corresponds to exactly $(c!)^N$ stub wiring configurations \cite{newman2018networks}: at each vertex, the $c$ incident edges can be assigned independently to the $c$ labelled stubs in $c!$ ways, and these assignments determine the corresponding stub pairing.\footnote{We could have defined our ensemble Markov chain microstates to be $(\boldsymbol{\sigma}, \mathbf{A})$ instead of $(\boldsymbol{\sigma}, \mathcal{C})$, where $\mathbf{A}$ is an $N \times N$ adjacency matrix consistent with all vertices having degree $c$. The reason is that each different adjacency matrix, i.e., each labelled graph, corresponds to the same number of stub wiring configurations, all producing isomorphic graphs.
We will continue with our definition of microstates as pairs $(\boldsymbol{\sigma}, \mathcal{C})$ because counting is more natural on stub wiring configurations, but the resulting equations would be identical with the alternative definition of $(\boldsymbol{\sigma}, \mathbf{A})$ pairs.} Configurations containing self-loops or multiple edges have fewer corresponding stub wirings because some such assignments are indistinguishable.
In the large-$N$ limit, however, the effect of self-loops and multiple edges is negligible, so we can focus only on simple graphs.

The total number of possible microstates of the Markov chain is

\begin{align}
\mathcal{N}_{\textrm{tot}} = 2^N \mathcal{N}_{N,c},
\nonumber
\end{align}

\noindent
where $\mathcal{N}_{N,c}$ is the number of different stub wiring configurations of $N$ vertices, each with $c$ stubs. In other words, $\mathcal{N}_{\textrm{tot}}$ is the total number of ways in which $N$ vertices can be coloured using two colours and their stubs ($c$ each) connected to form a graph.
Transitions in this Markov chain are only allowed between two microstates $(\boldsymbol{\sigma}^{[i]}, \mathcal{C}^{[j]})$ and $(\boldsymbol{\sigma}^{[k]}, \mathcal{C}^{[l]})$ if $j=l$, i.e., a change of vertex states on the same stub wiring configuration.
Thus there are $\mathcal{N}_{N,c}$ sets of microstates, corresponding to the number of different stub wirings, that are independent in the sense that if the system is in one of them at any time, it will stay there forever.
Thus the microscale Markov chain is reducible, but the advantage of this is that it makes the lumping combinatorics feasible.

Let us lump microstates together that have the same number of vertices and edges in each vertex and edge state. Note that this will lump together microstates that may have different stub wiring configurations, and hence are not reachable within the Markov chain, but this facilitates the combinatorics. A lumped state is again described by the vector $\mathbf{s} = ( N_{\textrm{I}}, N_{\textrm{S}}, N_{\textrm{II}}, N_{\textrm{SI}}, N_{\textrm{SS}})$. As we remarked earlier, this description is redundant (three of the five variables are sufficient to describe a lumped state), but for the sake of notational clarity, we continue using it. Each lumped state will contain microstates corresponding to many different ``colourings'' (assignments of one of the two possible vertex states to each vertex), and many different stub wiring configurations.

To apply our approach, we need to find the total number of microstates corresponding to the same lumped state $\mathbf{s}$, considering all possible arrangements $\boldsymbol{\sigma}$ of vertex states and stub wiring configurations $\mathcal{C}$.
This number is the product of the following factors: 
\begin{itemize}
\item The number of ways in which we can choose the $N_{\textrm{S}}$ susceptible vertices: $\binom{N}{N_{\textrm{S}}}$.
\item The number of ways in which we can choose the necessary number of stubs (emanating from susceptible vertices) to take part in $\textrm{SS}$-type edges: $\binom{cN_{\textrm{S}}}{2N_{\textrm{SS}}}$.
\item The number of ways in which we can make pairs of these stubs: $(2N_{\textrm{SS}})! / ( N_{\textrm{SS}}! \, 2^{N_{\textrm{SS}}} )$.
\item The number of ways in which we can choose the necessary number of stubs (emanating from infected vertices) to take part in $\textrm{II}$-type edges: $\binom{cN_{\textrm{I}}}{2N_{\textrm{II}}}$.
\item The number of ways in which we can make pairs of these stubs: $(2N_{\textrm{II}})! / ( N_{\textrm{II}}! \, 2^{N_{\textrm{II}}} )$.
\item The number of ways we can pair up the remaining stubs emanating from susceptible vertices with the (same number of) remaining stubs emanating from infected vertices to form the $\textrm{SI}$-type edges: $N_{\textrm{SI}}!$.
\end{itemize}
Putting all these parts together we have

\begin{align}
\mathcal{N}(\mathbf{s}) &= \binom{N}{N_{\textrm{S}}} \binom{cN_{\textrm{S}}}{2N_{\textrm{SS}}} \frac{(2N_{\textrm{SS}})!}{N_{\textrm{SS}}! \, 2^{N_{\textrm{SS}}}} \binom{cN_{\textrm{I}}}{2N_{\textrm{II}}} \frac{(2N_{\textrm{II}})!}{N_{\textrm{II}}! \, 2^{N_{\textrm{II}}}} N_{\textrm{SI}}!  \nonumber \\
&= \binom{N}{N_{\textrm{S}}} \binom{cN_{\textrm{S}}}{N_{\textrm{SI}}} \frac{(2N_{\textrm{SS}})!}{N_{\textrm{SS}}! \, 2^{N_{\textrm{SS}}}} \binom{cN_{\textrm{I}}}{N_{\textrm{SI}}} \frac{(2N_{\textrm{II}})!}{N_{\textrm{II}}! \, 2^{N_{\textrm{II}}}} N_{\textrm{SI}}!,
\label{eq:51.20}
\end{align}

\noindent
where we used $N_{\textrm{SI}} = cN_{\textrm{S}} - 2N_{\textrm{SS}} = cN_{\textrm{I}} - 2N_{\textrm{II}}$ to rewrite the binomial coefficients involving $N_{\textrm{SS}}$ and $N_{\textrm{II}}$.

To calculate the lumped transition rates $q_{ij}$ we must also calculate the number $\mathcal{N}^v (\mathcal{A}, \mathbf{d}, \mathbf{s})$ of microstates corresponding to $\mathbf{s}$ in which vertex $v$ has vertex-state $\mathcal{A}$ (either $\textrm{S}$ or $\textrm{I}$) and neighbourhood $\mathbf{d}=(d_{\textrm{I}},d_{\textrm{S}})$, where $d_{\textrm{S}}=c - d_{\textrm{I}}$. Let us first write $\mathcal{N}^v (\textrm{S}, \mathbf{d}, \mathbf{s})$. This number is the product of the following factors:
\begin{itemize}
\item The number of ways in which we can choose the $N_{\textrm{I}}$ infected vertices from all the remaining vertices: $\binom{N-1}{N_{\textrm{I}}}$.
\item The number of ways in which we can choose the $d_{\textrm{I}}$ stubs of vertex $v$ that are to be connected to infected vertices: $\binom{c}{d_{\textrm{I}}}$.
\item The number of ways we can choose the remaining $N_{\textrm{SI}}-d_{\textrm{I}}$ stubs, from all remaining stubs emanating from susceptible vertices, to be connected to infected vertices: $\binom{cN_{\textrm{S}}-c}{N_{\textrm{SI}}-d_{\textrm{I}}}$.
\item The number of ways we can choose the $N_{\textrm{SI}}$ stubs, emanating from infected vertices, to be connected to susceptible vertices: $\binom{cN_{\textrm{I}}}{N_{\textrm{SI}}}$.
\item The number of ways we can pair up the chosen $N_{\textrm{SI}}$ stubs emanating from susceptible vertices with the corresponding $N_{\textrm{SI}}$ stubs emanating from infected vertices: $N_{\textrm{SI}}!$.
\item The number of ways in which we can make pairs of the remaining stubs emanating from susceptible vertices, to be used in $\textrm{SS}$-type edges: $(2N_{\textrm{SS}})! / ( N_{\textrm{SS}}! \, 2^{N_{\textrm{SS}}} )$.
\item The number of ways in which we can make pairs of the remaining stubs emanating from infected vertices, to be used in $\textrm{II}$-type edges: $(2N_{\textrm{II}})! / ( N_{\textrm{II}}! \, 2^{N_{\textrm{II}}} )$.
\end{itemize}
Putting all these parts together we have

\begin{align}
\mathcal{N}^v (\textrm{S}, \mathbf{d}, \mathbf{s}) = \binom{N-1}{N_{\textrm{I}}} \binom{c}{d_{\textrm{I}}}  \binom{cN_{\textrm{S}}-c}{N_{\textrm{SI}}-d_{\textrm{I}}}  \binom{cN_{\textrm{I}}}{N_{\textrm{SI}}}  N_{\textrm{SI}}!  \frac{(2N_{\textrm{SS}})!}{N_{\textrm{SS}}! \, 2^{N_{\textrm{SS}}}}  \frac{(2N_{\textrm{II}})!}{N_{\textrm{II}}! \, 2^{N_{\textrm{II}}}},
\label{eq:51.31}
\end{align}

\noindent
and analogously

\begin{align}
\mathcal{N}^v (\textrm{I}, \mathbf{d}, \mathbf{s}) = \binom{N-1}{N_{\textrm{S}}} \binom{c}{c-d_{\textrm{I}}}  \binom{cN_{\textrm{I}}-c}{N_{\textrm{SI}}-c+d_{\textrm{I}}}  \binom{cN_{\textrm{S}}}{N_{\textrm{SI}}}  N_{\textrm{SI}}!  \frac{(2N_{\textrm{SS}})!}{N_{\textrm{SS}}! \, 2^{N_{\textrm{SS}}}}  \frac{(2N_{\textrm{II}})!}{N_{\textrm{II}}! \, 2^{N_{\textrm{II}}}}.
\label{eq:51.32}
\end{align}

The jump vectors describing the two possible transitions into other lumped states when the transition vertex has $d_{\textrm{I}}$ infected neighbours are now

\begin{align}
\boldsymbol{\Delta}_{d_{\textrm{I}}}^+ &= (1, -1, d_{\textrm{I}}, c-2d_{\textrm{I}}, d_{\textrm{I}}-c),  \nonumber \\ 
\boldsymbol{\Delta}_{d_{\textrm{I}}}^- &= (-1, 1, -d_{\textrm{I}}, 2d_{\textrm{I}}-c, c-d_{\textrm{I}}).  \nonumber 
\end{align}

\noindent
Using Eq. (\ref{eq:32.30}) we can now write the transition rate from lumped state $\mathbf{s}$ into lumped state $\mathbf{s} + \boldsymbol{\Delta}_{d_{\textrm{I}}}^+$, corresponding to an infection,

\begin{align}
q_{\mathbf{s}, \mathbf{s} + \boldsymbol{\Delta}_{d_{\textrm{I}}}^+} &= \frac{\alpha + d_{\textrm{I}} \beta}{ \mathcal{N}(\mathbf{s}) } \sum_{v \in V} \mathcal{N}^v (\textrm{S}, \mathbf{d}, \mathbf{s})   \nonumber \\
&= (\alpha + d_{\textrm{I}} \beta) N \frac{ \mathcal{N}^v (\textrm{S}, \mathbf{d}, \mathbf{s}) }{ \mathcal{N}(\mathbf{s}) }   \nonumber \\
&= (\alpha + d_{\textrm{I}} \beta) N_{\textrm{S}} \frac{ \binom{c}{d_{\textrm{I}}} \binom{cN_{\textrm{S}}-c}{N_{\textrm{SI}}-d_{\textrm{I}}} }{ \binom{cN_{\textrm{S}}}{N_{\textrm{SI}}} }   \nonumber \\
&= (\alpha + d_{\textrm{I}} \beta) N_{\textrm{S}} \,\, H(d_{\textrm{I}}; cN_{\textrm{S}},c,N_{\textrm{SI}}),
\label{eq:51.40}
\end{align}

\noindent
where we used the symmetry with respect to choosing the transition vertex $v$ and we used Eqs. (\ref{eq:51.31}), (\ref{eq:51.20}) and (\ref{eq:33.10}).
In completely analogous fashion we can write the transition rate from lumped state $\mathbf{s}$ into lumped state $\mathbf{s} + \boldsymbol{\Delta}_{d_{\textrm{I}}}^-$, corresponding to a recovery,

\begin{align}
q_{\mathbf{s}, \mathbf{s} + \boldsymbol{\Delta}_{d_{\textrm{I}}}^-} &= \frac{\gamma}{ \mathcal{N}(\mathbf{s}) } \sum_{v \in V} \mathcal{N}^v (\textrm{I}, \mathbf{d}, \mathbf{s})   \nonumber \\
&= \gamma N \frac{ \mathcal{N}^v (\textrm{I}, \mathbf{d}, \mathbf{s}) }{ \mathcal{N}(\mathbf{s}) }   \nonumber  \\
&= \gamma N_{\textrm{I}} \frac{ \binom{c}{c-d_{\textrm{I}}} \binom{cN_{\textrm{I}}-c}{N_{\textrm{SI}}-c+d_{\textrm{I}}} }{ \binom{cN_{\textrm{I}}}{N_{\textrm{SI}}} }   \nonumber \\
&= \gamma N_{\textrm{I}} \,\, H(c-d_{\textrm{I}}; cN_{\textrm{I}},c,N_{\textrm{SI}}).
\label{eq:51.60}
\end{align}

\noindent
The lumped transition rates, Eqs. (\ref{eq:51.40}) and (\ref{eq:51.60}), represent an approximation to the original dynamics that we defined on the ensemble of $(S, \mathcal{C})$ pairs.
One can verify that these rates are density-dependent, i.e., have the form of Eq. (\ref{eq:32.40}).
To obtain the ODEs describing the $N \to \infty$ limit of this approximate dynamics, we must write the function $\mathbf{F}(\mathbf{y})$, see Eq. (\ref{eq:32.60}),

\begin{align}
\mathbf{F} ( \mathbf{y} ) &= \sum_{d_{\textrm{I}}=0}^c \left[ \boldsymbol{\Delta}_{d_{\textrm{I}}}^+ \frac{   q_{\mathbf{s},\mathbf{s}+\boldsymbol{\Delta}_{d_{\textrm{I}}}^+}   }{ N } + \boldsymbol{\Delta}_{d_{\textrm{I}}}^- \frac{  q_{\mathbf{s},\mathbf{s}+\boldsymbol{\Delta}_{d_{\textrm{I}}}^-}  }{ N } \right]  \nonumber \\
&= \beta y_{\textrm{S}} \sum_{d_{\textrm{I}}=0}^c \boldsymbol{\Delta}_{d_{\textrm{I}}}^+ d_{\textrm{I}} H(d_{\textrm{I}}; cN_{\textrm{S}}, c, N_{\textrm{SI}}) + \alpha y_{\textrm{S}} \sum_{d_{\textrm{I}}=0}^c \boldsymbol{\Delta}_{d_{\textrm{I}}}^+ H(d_{\textrm{I}}; cN_{\textrm{S}}, c, N_{\textrm{SI}})  \nonumber \\
& \quad + \gamma y_{\textrm{I}} \sum_{d_{\textrm{I}}=0}^c \boldsymbol{\Delta}_{d_{\textrm{I}}}^- H(c-d_{\textrm{I}}; cN_{\textrm{I}}, c, N_{\textrm{SI}}).
\label{eq:51.70}
\end{align}

\noindent
As in the case of the cycle, due to the network being degree-regular, a single sum, over $d_{\textrm{I}}$, is sufficient to write $\mathbf{F} ( \mathbf{y} )$.
The sums in Eq. (\ref{eq:51.70}), for each component of the jump vectors, are averages over the hypergeometric distribution, involving only the first and second moments. Using Eqs. (\ref{eq:33.31}) and (\ref{eq:33.32}) we can write these averages directly.
The resulting differential equations, using Eq. (\ref{eq:32.50}), are

\begin{align}
\dot{y}_{\textrm{I}} &= \alpha y_{\textrm{S}} + \beta y_{\textrm{SI}} - \gamma y_{\textrm{I}}  \label{eq:51.221} \\
\dot{y}_{\textrm{S}} &= -\alpha y_{\textrm{S}} -\beta y_{\textrm{SI}} + \gamma y_{\textrm{I}} \label{eq:51.222} \\
\dot{y}_{\textrm{II}} &= \alpha y_{\textrm{SI}} + \beta \frac{c-1}{c} \frac{ y_{\textrm{SI}}^2 }{ y_{\textrm{S}} } +  \beta y_{\textrm{SI}} - 2 \gamma y_{\textrm{II}} \label{eq:51.223} \\
\dot{y}_{\textrm{SI}} &= \alpha (2 y_{\textrm{SS}} - y_{\textrm{SI}}) + \beta \frac{c-1}{c} \frac{ y_{\textrm{SI}} (2y_{\textrm{SS}}-y_{\textrm{SI}}) }{ y_{\textrm{S}} } - \beta y_{\textrm{SI}} - \gamma ( y_{\textrm{SI}} - 2 y_{\textrm{II}} ) \label{eq:51.224} \\
\dot{y}_{\textrm{SS}} &= -2 \alpha y_{\textrm{SS}} -2\beta \frac{c-1}{c} \frac{ y_{\textrm{SI}} y_{\textrm{SS}} }{ y_{\textrm{S}} } + \gamma y_{\textrm{SI}}. \label{eq:51.225}
\end{align}

\noindent
Keep in mind that $y_{\mathcal{AB}} = N_{\mathcal{AB}} / N$. If we prefer to use the---perhaps more natural---densities $e_{\mathcal{AB}} = N_{\mathcal{AB}} / L$, we must rescale by $2/c$, i.e.,

\begin{align}
e_{\mathcal{AB}} = \frac{2}{c} y_{\mathcal{AB}}.
\nonumber
\end{align}

\noindent
This results in the following set of equations,

\begin{align}
\dot{y}_{\textrm{I}} &= \alpha y_{\textrm{S}} + \beta \frac{c}{2} e_{\textrm{SI}} - \gamma y_{\textrm{I}}  \label{eq:51.241} \\
\dot{y}_{\textrm{S}} &= -\alpha y_{\textrm{S}} -\beta \frac{c}{2} e_{\textrm{SI}} + \gamma y_{\textrm{I}} \label{eq:51.242} \\
\dot{e}_{\textrm{II}} &= \alpha e_{\textrm{SI}} + \beta \frac{c-1}{2} \frac{ e_{\textrm{SI}}^2 }{ y_{\textrm{S}} } +  \beta e_{\textrm{SI}} - 2 \gamma e_{\textrm{II}} \label{eq:51.243} \\
\dot{e}_{\textrm{SI}} &= \alpha (2 e_{\textrm{SS}} - e_{\textrm{SI}}) + \beta \frac{c-1}{2} \frac{ e_{\textrm{SI}} (2e_{\textrm{SS}}-e_{\textrm{SI}}) }{ y_{\textrm{S}} } - \beta e_{\textrm{SI}} - \gamma ( e_{\textrm{SI}} - 2 e_{\textrm{II}} ) \label{eq:51.244} \\
\dot{e}_{\textrm{SS}} &= -2 \alpha e_{\textrm{SS}} -\beta (c-1) \frac{ e_{\textrm{SI}} e_{\textrm{SS}} }{ y_{\textrm{S}} } + \gamma e_{\textrm{SI}}. \label{eq:51.245}
\end{align}



\noindent
Eqs. (\ref{eq:51.221}--\ref{eq:51.225}) are equivalent to the homogeneous pairwise mean-field approximation in Eqs. (4.10a--4.10e) of Ref. \cite{kiss2017book}.
Note that Ref. \cite{kiss2017book} uses a convention different to ours. We define $N_{\textrm{II}}$, $N_{\textrm{SI}}$ and $N_{\textrm{SS}}$ to mean the number of edges connecting infected to infected vertices, susceptible to infected vertices and susceptible to susceptible vertices, respectively. In Ref. \cite{kiss2017book} the authors use the quantities $[\textrm{II}]$, $[\textrm{SI}] = [\textrm{IS}]$ and $[\textrm{SS}]$, which are edge counts in both directions, so they correspond to our quantities via $[\textrm{SI}] = N_{\textrm{SI}}$, $[\textrm{II}] = 2N_{\textrm{II}}$ and $[\textrm{SS}] = 2N_{\textrm{SS}}$.
Thus we were able to recover the homogeneous pairwise mean-field equations, for general degree.

\subsection{Erd\H{o}s-R\'enyi random graph ensemble}
\label{sec52}

Now we apply our approach to Erd\H{o}s-R\'enyi random networks.
Consider $N$ labelled vertices, where each vertex is in one of two vertex states, $\textrm{S}$ or $\textrm{I}$. The vertices are connected by exactly $L$ edges to form an undirected graph, and all the possible ways of doing this form the ensemble of graphs.
As was the case for random regular networks, here also some graphs will have self-loops and multiple edges. Our derivations will assume simple graphs and we note that the effect of self-loops and multiple edges becomes negligible in the large-$N$ limit.
Let us define a Markov chain whose microstates are specified by $(\boldsymbol{\sigma}, \mathbf{A})$, where $\boldsymbol{\sigma}$ is an $N$-tuple indicating the vertex state of each individual vertex, and $\mathbf{A}$ is an $N \times N$ adjacency matrix consistent with $L$ edges. The total number of possible microstates of this system is

\begin{align}
\mathcal{N}_{\textrm{tot}} = 2^N \mathcal{N}_{N,L},
\nonumber
\end{align}

\noindent
where $\mathcal{N}_{N,L}$ is the number of different $N \times N$ adjacency matrices consistent with $L$ edges. In other words, $\mathcal{N}_{\textrm{tot}}$ is the total number of ways in which $N$ vertices can be coloured using two colours and connected using $L$ edges to form a graph.
Transitions in this Markov chain are only allowed between two microstates $(\boldsymbol{\sigma}^{[i]}, \mathbf{A}^{[j]})$ and $(\boldsymbol{\sigma}^{[k]}, \mathbf{A}^{[l]})$ if $j=l$, i.e., a change of vertex states on the same graph. Thus there are $\mathcal{N}_{N,L}$ sets of microstates, corresponding to the number of different possible graph structures, that are independent, i.e., if the system is in one of them at any time, it will stay there forever.

As before, let us lump microstates together that have the same number of vertices and edges in each vertex and edge state. A lumped state is characterised by the vector $\mathbf{s} = ( N_{\textrm{I}}, N_{\textrm{S}}, N_{\textrm{II}}, N_{\textrm{SI}}, N_{\textrm{SS}} )$. Each lumped state will contain microstates corresponding to many different vertex state configurations and many different graph structures.
To apply our approach, we need to find the total number of microstates corresponding to the same lumped state $\mathbf{s}$,

\begin{align}
\mathcal{N}(\mathbf{s}) = \binom{N}{N_{\textrm{I}}} \binom{ \binom{N_{\textrm{I}}}{2} }{N_{\textrm{II}}} \binom{ \binom{N_{\textrm{S}}}{2} }{N_{\textrm{SS}}} \binom{N_{\textrm{I}} N_{\textrm{S}}}{N_{\textrm{SI}}}.
\label{eq:52.20}
\end{align}

\noindent
The first factor is the number of ways the $N_{\textrm{I}}$ infected vertices can be chosen from the set of all vertices. The second and third factors are the numbers of ways we can choose $N_{\textrm{II}}$ and $N_{\textrm{SS}}$ vertex pairs out of the possible $\binom{N_{\textrm{I}}}{2}$ and $\binom{N_{\textrm{S}}}{2}$, respectively, to use up all the $\textrm{II}$- and $\textrm{SS}$-type edges. The last factor is the number of ways we can choose $N_{\textrm{SI}}$ ``mixed'' vertex pairs out of the possible $N_{\textrm{I}} N_{\textrm{S}}$, to use up all the $\textrm{SI}$-type edges.\footnote{In the case of Erd\H{o}s-R\'enyi random networks we can perform the necessary counts by considering only connecting vertices and not stubs. Here it is not necessary to work at the level of stub wiring configurations, since vertex degrees have no constraint. In fact, for this exact reason, assigning the appropriate degrees first and then counting corresponding stub wiring configurations would be much more difficult in this case.}

To obtain the lumped transition rates $q_{ij}$ we must calculate the number $\mathcal{N}^v (\mathcal{A}, \mathbf{d}, \mathbf{s})$ of microstates corresponding to the lumped state $\mathbf{s}$ in which vertex $v$ has vertex state $\mathcal{A}$ (either $\textrm{I}$ or $\textrm{S}$) and neighbourhood $\mathbf{d}=(d_{\textrm{I}},d_{\textrm{S}})$. Similar reasoning as above gives us

\begin{align}
\mathcal{N}^v (\textrm{I}, \mathbf{d}, \mathbf{s}) = \binom{N-1}{N_{\textrm{I}}-1} \binom{N_{\textrm{I}}-1}{d_{\textrm{I}}} \binom{N_{\textrm{S}}}{d_{\textrm{S}}} \binom{ \binom{N_{\textrm{I}}-1}{2} }{N_{\textrm{II}}-d_{\textrm{I}}} \binom{ \binom{N_{\textrm{S}}}{2} }{N_{\textrm{SS}}} \binom{(N_{\textrm{I}}-1) N_{\textrm{S}}}{N_{\textrm{SI}}-d_{\textrm{S}}},
\label{eq:52.30}
\end{align}

\noindent
and

\begin{align}
\mathcal{N}^v (\textrm{S}, \mathbf{d}, \mathbf{s}) = \binom{N-1}{N_{\textrm{S}}-1} \binom{N_{\textrm{I}}}{d_{\textrm{I}}} \binom{N_{\textrm{S}}-1}{d_{\textrm{S}}} \binom{ \binom{N_{\textrm{I}}}{2} }{N_{\textrm{II}}} \binom{ \binom{N_{\textrm{S}}-1}{2} }{N_{\textrm{SS}}-d_{\textrm{S}}} \binom{N_{\textrm{I}}(N_{\textrm{S}}-1)}{N_{\textrm{SI}}-d_{\textrm{I}}}.
\label{eq:52.40}
\end{align}

\noindent
Let us explain the factors in Eq. (\ref{eq:52.30}); Eq. (\ref{eq:52.40}) is analogous. Vertex $v$ is fixed and is infected, so the first factor is the number of ways the remaining $N_{\textrm{I}}-1$ infected vertices can be chosen from the set of all remaining vertices. The second and third factors are the numbers of ways we can choose the $d_{\textrm{I}}$ infected and $d_{\textrm{S}}$ susceptible neighbours of vertex $v$ out of all the possible choices. The fourth and fifth factors are the numbers of ways we can assign $N_{\textrm{II}}-d_{\textrm{I}}$ edges of the $\textrm{II}$-type (note that we already used up $d_{\textrm{I}}$ such edges when assigning vertex $v$'s infected neighbours) and $N_{\textrm{SS}}$ edges of the $\textrm{SS}$ type. The last factor is the number of ways we can assign $N_{\textrm{SI}}-d_{\textrm{S}}$ edges of the $\textrm{SI}$ type, having already used up $d_{\textrm{S}}$ such edges when assigning vertex $v$'s susceptible neighbours.

Since vertex degrees are now not constrained, the jump vectors describing the two possible transitions into other lumped states when the transition vertex has $d_{\textrm{I}}$ infected and $d_{\textrm{S}}$ susceptible neighbours are the ones written for the general case, Eqs. (\ref{eq:32.10a}) and (\ref{eq:32.10b}).
Using Eq. (\ref{eq:32.30}) we can now write the transition rate from lumped state $\mathbf{s}$ into lumped state $\mathbf{s} + \boldsymbol{\Delta}_{d_{\textrm{I}},d_{\textrm{S}}}^+$, corresponding to an infection,

\begin{align}
q_{\mathbf{s},\mathbf{s} + \boldsymbol{\Delta}_{d_{\textrm{I}},d_{\textrm{S}}}^+} &= \frac{\alpha + d_{\textrm{I}} \beta}{ \mathcal{N}(\mathbf{s}) } \sum_{v \in V} \mathcal{N}^v (\textrm{S}, \mathbf{d}, \mathbf{s})   \nonumber \\
&= (\alpha + d_{\textrm{I}} \beta) N \frac{ \mathcal{N}^v (\textrm{S}, \mathbf{d}, \mathbf{s}) }{ \mathcal{N}(\mathbf{s}) }   \nonumber \\
&= (\alpha + d_{\textrm{I}} \beta) N_{\textrm{S}} \frac{ \binom{N_{\textrm{I}}}{d_{\textrm{I}}} \binom{N_{\textrm{S}}-1}{d_{\textrm{S}}} \binom{ \binom{N_{\textrm{S}}-1}{2} }{N_{\textrm{SS}}-d_{\textrm{S}}} \binom{N_{\textrm{I}}(N_{\textrm{S}}-1)}{N_{\textrm{SI}}-d_{\textrm{I}}} }{ \binom{ \binom{N_{\textrm{S}}}{2} }{N_{\textrm{SS}}} \binom{N_{\textrm{I}} N_{\textrm{S}}}{N_{\textrm{SI}}} }  \nonumber \\
&= (\alpha + d_{\textrm{I}} \beta) N_{\textrm{S}} \,\, H(d_{\textrm{I}}; N_{\textrm{I}} N_{\textrm{S}}, N_{\textrm{I}}, N_{\textrm{SI}}) \,\, H \left( d_{\textrm{S}}; \binom{N_{\textrm{S}}}{2}, N_{\textrm{S}}-1, N_{\textrm{SS}} \right),
\label{eq:52.50}
\end{align}

\noindent
where we used the symmetry with respect to choosing the transition vertex $v$ and we used Eqs. (\ref{eq:52.40}), (\ref{eq:52.20}) and (\ref{eq:33.10}).
In completely analogous fashion we can write the transition rate from lumped state $\mathbf{s}$ into lumped state $\mathbf{s} + \boldsymbol{\Delta}_{d_{\textrm{I}},d_{\textrm{S}}}^-$, corresponding to a recovery,

\begin{align}
q_{\mathbf{s}, \mathbf{s} + \boldsymbol{\Delta}_{d_{\textrm{I}},d_{\textrm{S}}}^-} &= \frac{\gamma}{ \mathcal{N}(\mathbf{s}) } \sum_{v \in V} \mathcal{N}^v (\textrm{I}, \mathbf{d}, \mathbf{s})   \nonumber \\
&= \gamma N \frac{ \mathcal{N}^v (\textrm{I}, \mathbf{d}, \mathbf{s}) }{ \mathcal{N}(\mathbf{s}) }   \nonumber \\
&= \gamma N_{\textrm{I}} \frac{ \binom{N_{\textrm{I}}-1}{d_{\textrm{I}}} \binom{N_{\textrm{S}}}{d_{\textrm{S}}} \binom{ \binom{N_{\textrm{I}}-1}{2} }{N_{\textrm{II}}-d_{\textrm{I}}} \binom{(N_{\textrm{I}}-1) N_{\textrm{S}}}{N_{\textrm{SI}}-d_{\textrm{S}}} }{ \binom{ \binom{N_{\textrm{I}}}{2} }{N_{\textrm{II}}} \binom{N_{\textrm{I}} N_{\textrm{S}}}{N_{\textrm{SI}}} }  \nonumber \\
&= \gamma N_{\textrm{I}} \,\, H \left( d_{\textrm{I}}; \binom{N_{\textrm{I}}}{2}, N_{\textrm{I}}-1, N_{\textrm{II}} \right) \,\, H( d_{\textrm{S}}; N_{\textrm{S}} N_{\textrm{I}}, N_{\textrm{S}}, N_{\textrm{SI}}).
\label{eq:52.60}
\end{align}

\noindent
The lumped transition rates, Eqs. (\ref{eq:52.50}) and (\ref{eq:52.60}), represent an approximation to the original dynamics that we defined on the ensemble of $(S, \mathbf{A})$ pairs.
One can verify that these rates are density-dependent, i.e., have the form of Eq. (\ref{eq:32.40}).
To obtain the ODEs describing the $N \to \infty$ limit of this approximate dynamics, we must write the function $\mathbf{F}(\mathbf{y})$, see Eq. (\ref{eq:32.60}),

\begin{align}
\mathbf{F}(\mathbf{y}) = \boldsymbol{\Phi}^+(\mathbf{y}) + \boldsymbol{\Phi}^-(\mathbf{y}),
\label{eq:52.80}
\end{align}

\noindent
where

\begin{align}
\boldsymbol{\Phi}^+(\mathbf{y}) &= \sum_{d_{\textrm{I}}=0}^{ \min(N_{\textrm{SI}}, N_{\textrm{I}}) } \,\, \sum_{d_{\textrm{S}}=0}^{ \min(N_{\textrm{SS}}, N_{\textrm{S}}-1) } \boldsymbol{\Delta}_{d_{\textrm{I}},d_{\textrm{S}}}^+    \frac{   q_{\mathbf{s}, \mathbf{s} + \boldsymbol{\Delta}_{d_{\textrm{I}},d_{\textrm{S}}}^+}   }{N},      \label{eq:52.91}   \\
\boldsymbol{\Phi}^-(\mathbf{y}) &= \sum_{d_{\textrm{I}}=0}^{ \min(N_{\textrm{II}}, N_{\textrm{I}}-1) } \,\, \sum_{d_{\textrm{S}}=0}^{ \min(N_{\textrm{SI}}, N_{\textrm{S}}) } \boldsymbol{\Delta}_{d_{\textrm{I}},d_{\textrm{S}}}^-    \frac{   q_{\mathbf{s}, \mathbf{s} + \boldsymbol{\Delta}_{d_{\textrm{I}},d_{\textrm{S}}}^-}   }{N}.      \label{eq:52.92}
\end{align}

\noindent
Note that the upper limits on the sums are the highest possible values of $d_{\textrm{I}}$ and $d_{\textrm{S}}$ in our construction.
The right-hand sides of Eqs. (\ref{eq:52.91}) and (\ref{eq:52.92}) are functions of $\mathbf{y}$ via $\mathbf{s} = N \mathbf{y}$.
The functions $\boldsymbol{\Phi}^+(\mathbf{y})$ and $\boldsymbol{\Phi}^-(\mathbf{y})$, written out in full, are

\begin{align}
\boldsymbol{\Phi}^+(\mathbf{y}) &= \beta y_{\textrm{S}} \sum_{d_{\textrm{I}}=0}^{ \min(N_{\textrm{SI}}, N_{\textrm{I}}) } \,\, \sum_{d_{\textrm{S}}=0}^{ \min(N_{\textrm{SS}}, N_{\textrm{S}}-1) } \boldsymbol{\Delta}_{d_{\textrm{I}},d_{\textrm{S}}}^+ d_{\textrm{I}} \,\,  H(d_{\textrm{I}}; N_{\textrm{I}} N_{\textrm{S}}, N_{\textrm{I}}, N_{\textrm{SI}})   \nonumber \\
& \quad \times H \left( d_{\textrm{S}}; \binom{N_{\textrm{S}}}{2}, N_{\textrm{S}}-1, N_{\textrm{SS}} \right)  \nonumber \\
& + \alpha y_{\textrm{S}} \sum_{d_{\textrm{I}}=0}^{ \min(N_{\textrm{SI}}, N_{\textrm{I}}) } \,\, \sum_{d_{\textrm{S}}=0}^{ \min(N_{\textrm{SS}}, N_{\textrm{S}}-1) } \boldsymbol{\Delta}_{d_{\textrm{I}},d_{\textrm{S}}}^+ \,\,  H(d_{\textrm{I}}; N_{\textrm{I}} N_{\textrm{S}}, N_{\textrm{I}}, N_{\textrm{SI}})   \nonumber \\
& \quad \times H \left( d_{\textrm{S}}; \binom{N_{\textrm{S}}}{2}, N_{\textrm{S}}-1, N_{\textrm{SS}} \right),
\label{eq:52.100}
\end{align}

\noindent
and

\begin{align}
\boldsymbol{\Phi}^-(\mathbf{y}) &= \gamma y_{\textrm{I}} \sum_{d_{\textrm{I}}=0}^{ \min(N_{\textrm{II}}, N_{\textrm{I}}-1) } \,\, \sum_{d_{\textrm{S}}=0}^{ \min(N_{\textrm{SI}}, N_{\textrm{S}}) } \boldsymbol{\Delta}_{d_{\textrm{I}},d_{\textrm{S}}}^- \,\,  H \left( d_{\textrm{I}}; \binom{N_{\textrm{I}}}{2}, N_{\textrm{I}}-1, N_{\textrm{II}} \right)  \nonumber \\
& \quad \times H \left( d_{\textrm{S}}; N_{\textrm{S}} N_{\textrm{I}}, N_{\textrm{S}}, N_{\textrm{SI}} \right).      \label{eq:52.110}
\end{align}

\noindent
The double sums in Eqs. (\ref{eq:52.100}) and (\ref{eq:52.110}) factorise into products of averages over the hypergeometric distribution. As before, these averages involve only the first and second moments, and using Eqs. (\ref{eq:33.31}) and (\ref{eq:33.32}) we can write them directly.
The resulting differential equations, using Eq. (\ref{eq:32.50}), are

\begin{align}
\dot{y}_{\textrm{I}} &= \alpha y_{\textrm{S}} + \beta y_{\textrm{SI}} - \gamma y_{\textrm{I}}  \label{eq:52.271} \\
\dot{y}_{\textrm{S}} &= -\alpha y_{\textrm{S}} -\beta y_{\textrm{SI}} + \gamma y_{\textrm{I}} \label{eq:52.272} \\
\dot{y}_{\textrm{II}} &= \alpha y_{\textrm{SI}} + \beta \left( \frac{ y_{\textrm{SI}}^2 }{ y_{\textrm{S}} } + y_{\textrm{SI}} \right) - 2 \gamma y_{\textrm{II}} \label{eq:52.273} \\
\dot{y}_{\textrm{SI}} &= \alpha (2 y_{\textrm{SS}} - y_{\textrm{SI}}) + \beta \left( \frac{ 2 y_{\textrm{SI}} y_{\textrm{SS}} }{ y_{\textrm{S}} } -\frac{ y_{\textrm{SI}}^2 }{ y_{\textrm{S}} } - y_{\textrm{SI}} \right) - \gamma ( y_{\textrm{SI}} - 2 y_{\textrm{II}} ) \label{eq:52.274} \\
\dot{y}_{\textrm{SS}} &= -2 \alpha y_{\textrm{SS}} -2\beta \frac{ y_{\textrm{SI}} y_{\textrm{SS}} }{ y_{\textrm{S}} } + \gamma y_{\textrm{SI}}. \label{eq:52.275}
\end{align}

\noindent
Keep in mind that $y_{\mathcal{AB}} = N_{\mathcal{AB}} / N$. If we prefer to use the densities $e_{\mathcal{AB}} = N_{\mathcal{AB}} / L$, we must rescale by $2/z$, i.e.,

\begin{align}
e_{\mathcal{AB}} = \frac{2}{z} y_{\mathcal{AB}},
\nonumber
\end{align}

\noindent
where $z = 2L/N$ is the mean degree. This results in the following set of equations,

\begin{align}
\dot{y}_{\textrm{I}} &= \alpha y_{\textrm{S}} + \beta \frac{z}{2} e_{\textrm{SI}} - \gamma y_{\textrm{I}}  \label{eq:52.281} \\
\dot{y}_{\textrm{S}} &= -\alpha y_{\textrm{S}} -\beta \frac{z}{2} e_{\textrm{SI}} + \gamma y_{\textrm{I}} \label{eq:52.282} \\
\dot{e}_{\textrm{II}} &= \alpha e_{\textrm{SI}} + \beta \left( \frac{z}{2} \frac{ e_{\textrm{SI}}^2 }{ y_{\textrm{S}} } + e_{\textrm{SI}} \right) - 2 \gamma e_{\textrm{II}} \label{eq:52.283} \\
\dot{e}_{\textrm{SI}} &= \alpha (2 e_{\textrm{SS}} - e_{\textrm{SI}}) + \beta \left( z \frac{ e_{\textrm{SI}} e_{\textrm{SS}} }{ y_{\textrm{S}} } - \frac{z}{2} \frac{ e_{\textrm{SI}}^2 }{ y_{\textrm{S}} } - e_{\textrm{SI}} \right) - \gamma ( e_{\textrm{SI}} - 2 e_{\textrm{II}} ) \label{eq:52.284} \\
\dot{e}_{\textrm{SS}} &= -2 \alpha e_{\textrm{SS}} -\beta z \frac{ e_{\textrm{SI}} e_{\textrm{SS}} }{ y_{\textrm{S}} } + \gamma e_{\textrm{SI}}, \label{eq:52.285}
\end{align}

\noindent
and the mean degree already appears explicitly in the equations.
Eqs. (\ref{eq:52.271}--\ref{eq:52.275}) have the same structure as the homogeneous pairwise mean-field approximation in Eqs. (4.10a--4.10e) of Ref. \cite{kiss2017book}, but without a $(n-1)/n$ prefactor for the quadratic terms in Eqs. (\ref{eq:52.273}--\ref{eq:52.275}).
Note again the difference in edge count convention: the quantities $[\textrm{SI}]$, $[\textrm{II}]$, $[\textrm{SS}]$ in Ref. \cite{kiss2017book} are related to our edge counts via $[\textrm{SI}] = N_{\textrm{SI}}$, $[\textrm{II}] = 2N_{\textrm{II}}$ and $[\textrm{SS}] = 2N_{\textrm{SS}}$.
The prefactor $\kappa = 1$, instead of $\kappa = (n-1)/n$, appears in the standard pairwise closure on configuration model networks with a Poisson degree distribution---which are identical to Erd\H{o}s-R\'enyi networks in the large-$N$ limit---see, e.g., Ref. \cite{kiss2023necessary}. Our approach provides an independent path to arrive at these ODEs, focusing solely on the explicit averaging of transition rates between microstates, without invoking any moment closure arguments.

\section{Discussion and conclusions}
\label{sec6}

Building on previous work \cite{ward2022micro, ward2025mean}, in this paper we showed how approximate lumping of Markov chains may be used to derive EBMF approximations for a broad class of dynamical processes on networks. Our approach relies on a natural refinement of vertex-state-based population models \cite{ward2022micro}: we define lumping partitions where each microstate has the same number of vertices in each vertex state, but also the same number of edges in each edge state. Transition rates between microstates in different lumped states are averaged to obtain a reduced, approximate Markov chain description of the system. This approximation step reduces the size of the Markov chain state space from exponential to polynomial in system size $N$. The resulting lumped Markov chain is generally density-dependent, which behaves deterministically in the $N \to \infty$ limit, governed by a low-dimensional set of ODEs.
An important feature of our framework is that the procedure follows in a fully systematic way, and applies uniformly to both single graphs and graph ensembles. The only challenge lies in performing the microstate counts, required to determine the transition rates between lumped states; aside from this step, the derivation proceeds in a general and network-structure-independent manner.

We demonstrated our approach for a selection of networks where the combinatorial counting can be performed easily: cycles, random regular and Erd\H{o}s-R\'enyi graph ensembles. Our results show that well-known EBMF approximations arise as special cases of our general approach \cite{kiss2017book}. In particular, for the examples considered, the resulting equations coincide with classical pairwise mean-field models derived using moment closure arguments. This provides a new perspective on such models, placing them within a principled mathematical framework and clarifying the precise nature of the averaging implicit in their derivation.

The question of whether the lumped transition rates are always density-dependent is an interesting one. The combinatorial calculations required to establish this property depend only on the underlying network structure. While our intuition suggests that density dependence always holds for the class of models considered here, we have not established this result for general networks, and it therefore remains an open problem for future work.
In the examples studied in this paper, the lumped transition rates could be expressed compactly in terms of the hypergeometric distribution.
We conjecture that for network structures and lumping partitions where the combinatorial calculations can be expressed in this form, the lumped transition rates will be density-dependent and the large-$N$ limit ODEs can be written as averages over the hypergeometric distribution.
In particular, within the affine model class considered here, only the first two moments of the hypergeometric distribution are required.
These observations may open the door to applying approximate lumping to a much broader class of dynamical models, which we also leave for future work.

Our results for graph ensembles highlight the flexibility of the approximate lumping framework. The microstate-counting approach developed in this setting appears, in principle, to extend to more structured graph ensembles, including configuration-model ensembles and related refinements, although the associated combinatorial calculations are likely to become increasingly involved. This suggests a natural route toward a hierarchy of lumping-based approximations, corresponding to progressively more detailed network ensemble descriptions; see Ref. \cite{khudabukhsh2019approximate} for related work.

We presented our approach through the example of the binary-state SISa model for simplicity. All of our results are also straightforward to write for the general model of Sec. \ref{sec21},
however the exposition becomes rather complex and not particularly insightful.
We thus suggest tackling the derivation of EBMF approximations for more complex models on a case-by-case basis. Alternatively, we imagine that a computational algebra approach could be employed to automate this process.

Overall, our Markov chain lumping framework provides a principled mathematical foundation for mean-field approximations on networks, making explicit the averaging procedure that underlies them and identifying the origin of the resulting approximation error. This perspective offers a promising route towards quantifying that error, an important and challenging open problem.

\section*{Funding}

J.A. Ward and G. Tim\'ar acknowledge funding from the Leverhulme Trust Project Grant number RPG-2023-187. P.L. Simon acknowledges support from the Hungarian Scientific Research Fund, OTKA (grant no. 135241) and from ERC Synergy Grant No. 810115 - DYNASNET.

\begin{appendices}

\renewcommand{\theequation}{\Alph{section}\arabic{equation}}
\renewcommand{\theHequation}{\Alph{section}.\arabic{equation}}
\setcounter{equation}{0}

\section{Counting microstates on a cycle}
\label{secA}

In this Appendix we derive two results used in the main text: the number of microstates of a cycle corresponding to a given $(N_{\textrm{I}}, N_{\textrm{SI}})$ and corresponding to a given $(\mathcal{A}, d_{\textrm{I}}, N_{\textrm{I}}, N_{\textrm{SI}})$. We may think of any given microstate as a colouring of the cycle, where infected vertices are coloured red and susceptible ones are coloured blue. This terminology is used throughout this Appendix, as it will make discussion easier.

For the derivations it will be useful to consider \emph{clusters} of red and blue vertices along the cycle. By clusters we mean uninterrupted sequences of vertices of the same colour. When $0 < N_{\textrm{I}} < N$, the number $\chi$ of clusters (of either red or blue vertices) is simply $N_{\textrm{SI}} / 2$, so we may also use $(N_{\textrm{I}}, \chi)$ to identify lumped states. In the cases $N_{\textrm{I}}=0$ and $N_{\textrm{I}}=N$, we have $N_{\textrm{SI}}=0$ but clearly one single cluster of either blue ($N_{\textrm{I}}=0$) or red ($N_{\textrm{I}}=N$) vertices, so we must deal with these boundary cases separately. In what follows we will work with $\chi$ where possible, but write the final results in terms of $N_{\textrm{SI}}$.

\subsection{Number of colourings of a cycle that satisfy $(N_{\textrm{I}}, N_{\textrm{SI}})$}
\label{secA1}

\begin{figure}[H]
\centering
\includegraphics[width=0.7\textwidth]{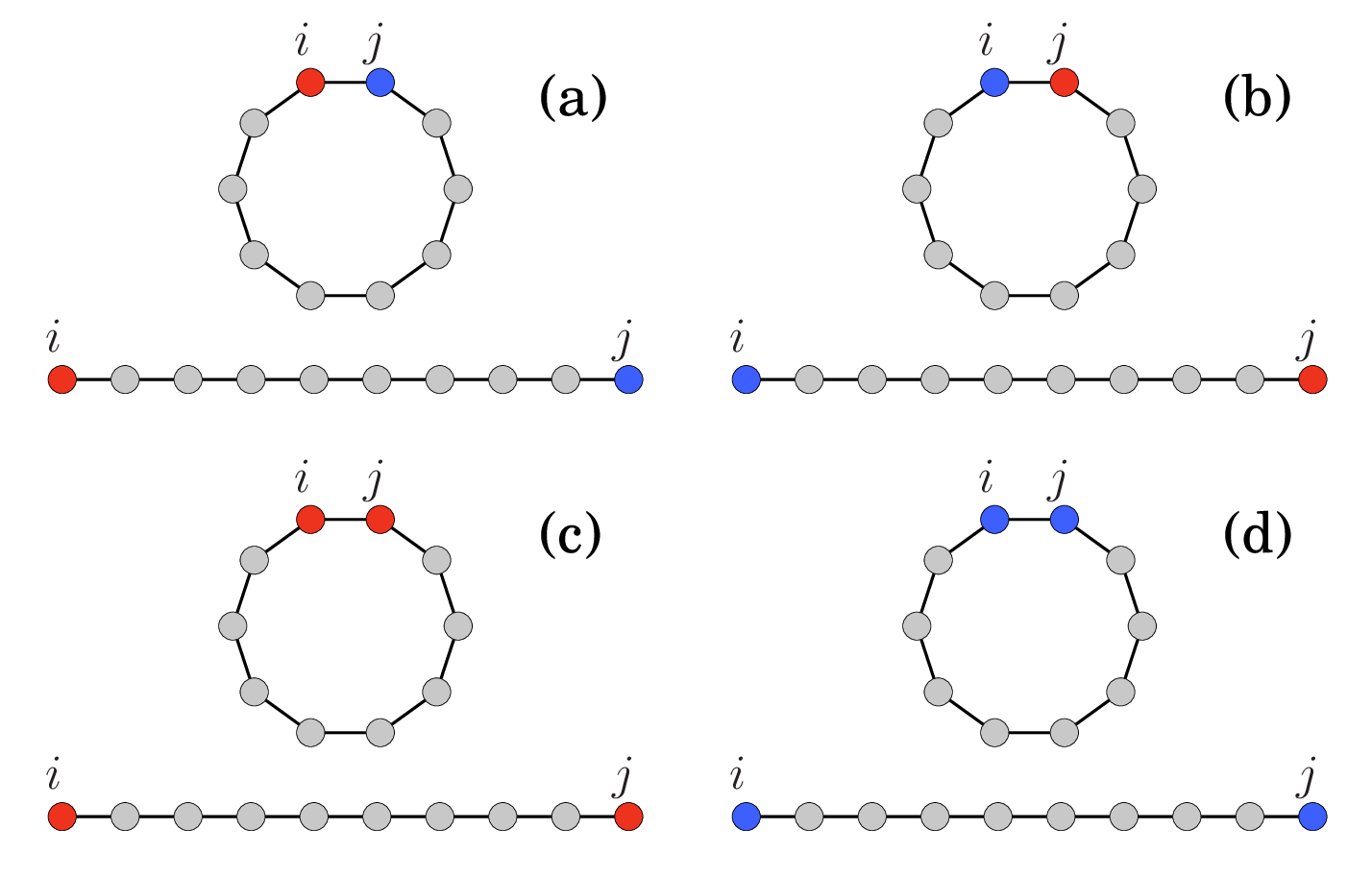}
\caption{Four possible ways of colouring two adjacent labelled vertices on a cycle. Grey vertices mean variable colours. In all four panels the given cycle is also shown stretched out into a line, starting at vertex $i$, ending at vertex $j$.}
\label{fig:full_cycle_fig}
\end{figure}

To calculate $\mathcal{N}(\mathbf{s})$ we separate the set of all colourings of a cycle that satisfy $(N_{\textrm{I}}, N_{\textrm{SI}})$ [or equivalently, that satisfy $(N_{\textrm{I}}, \chi)$ when $0 < N_{\textrm{I}} < N$] into four cases, according to the four possible colourings of a pair of adjacent labelled vertices, depicted in Fig. \ref{fig:full_cycle_fig}. If, as in Fig. \ref{fig:full_cycle_fig}(a), two adjacent vertices, $i$ and $j$, are coloured red and blue, the number of colourings of the rest of the cycle---that satisfy $(N_{\textrm{I}}, \chi)$ for the entire cycle---is the same as the number of possible red cluster size sequences, multiplied by the number of possible blue cluster size sequences along the line $i$---$j$ [see Fig. \ref{fig:full_cycle_fig}(a)], where there are $N_{\textrm{I}}$ red and $N_{\textrm{S}} = N - N_{\textrm{I}}$ blue vertices and $\chi$ clusters of each colour. (Vertex $i$ belongs to the first red cluster and vertex $j$ belongs to the last blue cluster along the line.) Since each cluster must contain at least one vertex, this number is

\begin{align}
\mathcal{N}_{\textrm{a}} = \binom{N_{\textrm{I}} - 1}{\chi-1} \binom{N_{\textrm{S}} - 1}{\chi-1},
\label{eq:A1.10}
\end{align}

\noindent
for all realisable values of $N_{\textrm{I}}$ and $\chi$.
By symmetry, the number corresponding to Fig. \ref{fig:full_cycle_fig}(b) is the same,

\begin{align}
\mathcal{N}_{\textrm{b}} = \binom{N_{\textrm{S}} - 1}{\chi-1} \binom{N_{\textrm{I}} - 1}{\chi-1},
\label{eq:A1.20}
\end{align}

\noindent
for all realisable values of $N_{\textrm{I}}$ and $\chi$.

\noindent
When, as in Fig. \ref{fig:full_cycle_fig}(c), both vertices $i$ and $j$ are red, the number of colourings of the rest of the cycle is equal to the number of possible
red cluster size sequences, multiplied by the number of possible blue cluster size sequences along the line $i$---$j$, where there are $N_{\textrm{I}}$ red vertices in $\chi+1$ clusters and $N_{\textrm{S}}=N-N_{\textrm{I}}$ blue vertices in $\chi$ clusters. (We must consider $\chi+1$ red clusters along the line $i$---$j$, because vertices $i$ and $j$, while belonging to separate clusters on the line, belong to the same cluster on the cycle.) The number of possible colourings for case (c) are thus

\begin{align}
\mathcal{N}_{\textrm{c}} =
\begin{cases}
\binom{N_{\textrm{I}} - 1}{\chi} \binom{N_{\textrm{S}} - 1}{\chi-1} \quad & 2 \leq N_{\textrm{I}} \leq N-1 \\
1 \quad & N_{\textrm{I}}=N,
\end{cases}
\label{eq:A1.30}
\end{align}

\noindent
for all realisable values of $\chi$.
By symmetry, for case (d) we have,

\begin{align}
\mathcal{N}_{\textrm{d}} =
\begin{cases}
\binom{N_{\textrm{S}} - 1}{\chi} \binom{N_{\textrm{I}} - 1}{\chi-1} \quad & 1 \leq N_{\textrm{I}} \leq N-2 \\
1 \quad & N_{\textrm{I}}=0,
\end{cases}
\label{eq:A1.40}
\end{align}

\noindent
for all realisable values of $\chi$.
In total the number of possible colourings of the cycle that satisfy $(N_{\textrm{I}}, \chi)$ is

\begin{align}
\mathcal{N}(\mathbf{s}) = \mathcal{N}_{\textrm{a}} + \mathcal{N}_{\textrm{b}} + \mathcal{N}_{\textrm{c}} + \mathcal{N}_{\textrm{d}}.
\label{eq:A1.50}
\end{align}

\noindent
Using Eqs. (\ref{eq:A1.10}--\ref{eq:A1.50}), expanding the binomials and simplifying, we get

\begin{align}
\mathcal{N}(\mathbf{s}) =
\begin{cases}
\frac{N N_{\textrm{SI}}}{2 N_{\textrm{I}} N_{\textrm{S}}} \binom{N_{\textrm{I}}}{N_{\textrm{SI}}/2} \binom{N_{\textrm{S}}}{N_{\textrm{SI}}/2} \quad & 1 \leq N_{\textrm{I}} \leq N-1 \\
1 \quad & N_{\textrm{I}}=0 \quad \text{or} \quad N_{\textrm{I}}=N,
\end{cases}
\label{eq:A1.60}
\end{align}

\noindent
for all realisable values of $N_{\textrm{SI}}$.

\subsection{Number of colourings that satisfy $(\mathcal{A}, d_{\textrm{I}}, N_{\textrm{I}}, N_{\textrm{SI}})$}
\label{secA2}

\begin{figure}[H]
\centering
\includegraphics[width=0.7\textwidth]{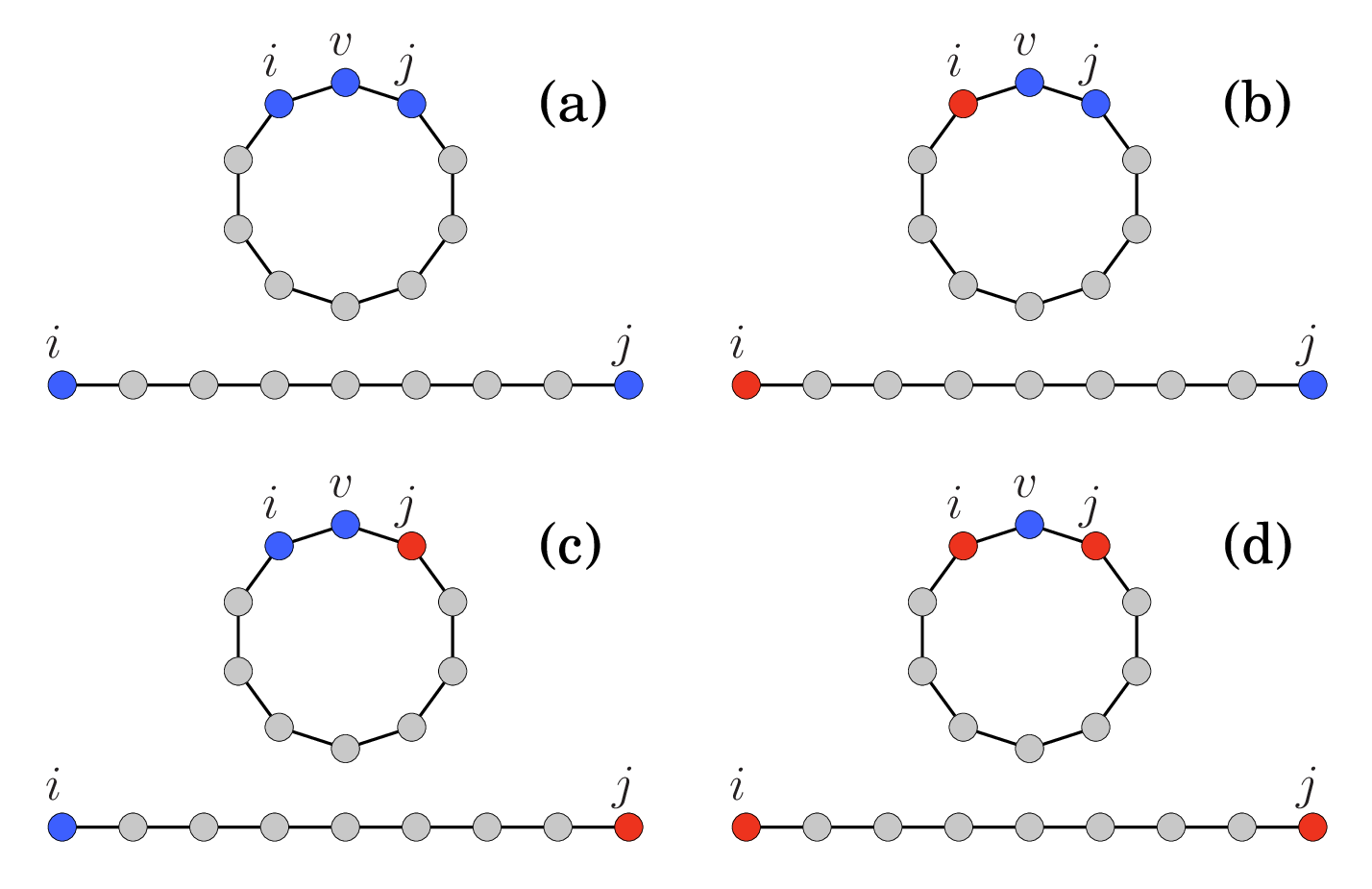}
\caption{Four possible ways of colouring vertices $v,i,j$ when vertex $v$ is susceptible. Grey vertices mean variable colours. In all four panels the given cycle is also shown stretched out into a line (excluding vertex $v$), starting at vertex $i$, ending at vertex $j$.}
\label{fig:v_cycle_fig_1}
\end{figure}

Similarly to the calculation of $\mathcal{N}(\mathbf{s})$ we separate the colourings that satisfy $(\mathcal{A}, d_{\textrm{I}}, N_{\textrm{I}}, N_{\textrm{SI}})$ into groups according to $d_{\textrm{I}}$. We start by counting colourings where a reference vertex $v$ is in state $\textrm{S}$. If both neighbours of this vertex are also in state $\textrm{S}$, corresponding to $d_{\textrm{I}}=0$, depicted in Fig. \ref{fig:v_cycle_fig_1}(a), then the number of colourings is given by the number of possible blue cluster size sequences, multiplied by the number of possible red cluster size sequences along the line $i$---$j$, where there are $N_{\textrm{S}}-1$ blue vertices (vertex $v$ is discounted) in $\chi+1$ clusters and $N_{\textrm{I}}$ red vertices in $\chi$ clusters. (We must consider $\chi+1$ blue clusters along the line $i$---$j$, because vertices $i$ and $j$ belong to the same cluster on the cycle.)
This number is

\begin{align}
\mathcal{N}_{\textrm{a}} = \mathcal{N}(\textrm{S}, 0, N_{\textrm{I}}, N_{\textrm{SI}}) =
\begin{cases}
\binom{N_{\textrm{S}}-2}{\chi} \binom{N_{\textrm{I}}-1}{\chi-1} \quad & 1 \leq N_{\textrm{I}} \leq N-3 \\
1 \quad & N_{\textrm{I}} = 0,
\end{cases}
\label{eq:A2.10}
\end{align}

\noindent
for all realisable values of $\chi$, and we recall that $\chi = N_{\textrm{SI}}/2$ when $N_{\textrm{SI}}>0$.

\noindent
If one neighbour of vertex $v$ is infected ($d_{\textrm{I}}=1$), depicted in Fig. \ref{fig:v_cycle_fig_1}(b,c), then the number of colourings is given by the number of possible
red cluster size sequences, multiplied by the number of possible blue cluster size sequences along the line $i$---$j$, where there are $N_{\textrm{I}}$ red vertices in $\chi$ clusters and $N_{\textrm{S}}-1$ blue vertices (vertex $v$ is discounted) in $\chi$ clusters. We get

\begin{align}
\mathcal{N}_{\textrm{b}} = \binom{N_{\textrm{I}} - 1}{\chi-1} \binom{N_{\textrm{S}} - 2}{\chi-1},
\label{eq:A2.20}
\end{align}

\noindent
for all realisable values of $N_{\textrm{I}}$ and $\chi$.
By symmetry,

\begin{align}
\mathcal{N}_{\textrm{c}} = \binom{N_{\textrm{S}} - 2}{\chi-1} \binom{N_{\textrm{I}} - 1}{\chi-1},
\label{eq:A2.30}
\end{align}

\noindent
for all realisable values of $N_{\textrm{I}}$ and $\chi$, so we have

\begin{align}
\mathcal{N}_{\textrm{b}} + \mathcal{N}_{\textrm{c}} = \mathcal{N}(\textrm{S}, 1, N_{\textrm{I}}, N_{\textrm{SI}}) = 2 \binom{N_{\textrm{I}}-1}{\chi-1} \binom{N_{\textrm{S}}-2}{\chi-1},
\label{eq:A2.40}
\end{align}

\noindent
for all realisable values of $N_{\textrm{I}}$ and $\chi$.
Finally, in the case where both neighbours of vertex $v$ are infected ($d_{\textrm{I}}=2$), depicted in Fig. \ref{fig:v_cycle_fig_1}(d), the number of colourings is given by the number of possible
red cluster size sequences, multiplied by the number of possible blue cluster size sequences along the line $i$---$j$, where there are $N_{\textrm{I}}$ red vertices in $\chi$ clusters and $N_{\textrm{S}}-1$ blue vertices (vertex $v$ is discounted) in $\chi-1$ clusters (the isolated blue vertex $v$ is a cluster in itself). This number is

\begin{align}
\mathcal{N}_{\textrm{d}} = \mathcal{N}(\textrm{S}, 2, N_{\textrm{I}}, N_{\textrm{SI}}) =
\begin{cases}
\binom{N_{\textrm{I}}-1}{\chi-1} \binom{N_{\textrm{S}}-2}{\chi-2} \quad & 2 \leq N_{\textrm{I}} \leq N-2 \\
1 \quad & N_{\textrm{I}}=N-1,
\end{cases}
\label{eq:A2.50}
\end{align}

\noindent
for all realisable values of $\chi$.
Equations (\ref{eq:A2.10}), (\ref{eq:A2.40}) and (\ref{eq:A2.50}) can be unified as

\begin{align}
\mathcal{N}^v(\textrm{S}, \mathbf{d}, \mathbf{s}) =
\begin{cases}
\binom{N_{\textrm{I}}-1}{ N_{\textrm{SI}}/2 -1} \binom{2}{d_{\textrm{I}}} \binom{N_{\textrm{S}}-2}{N_{\textrm{SI}}/2-d_{\textrm{I}}} \quad & 1 \leq N_{\textrm{I}} \leq N-2 \\
1 \quad & N_{\textrm{I}} = 0 \quad \text{or} \quad N_{\textrm{I}}=N-1,
\end{cases}
\label{eq:A2.60}
\end{align}

\noindent
for all realisable values of $d_{\textrm{I}}$ and $N_{\textrm{SI}}$.
The four cases where the reference vertex $v$ is infected are completely analogous, so we can immediately write

\begin{align}
\mathcal{N}^v(\textrm{I}, \mathbf{d}, \mathbf{s}) =
\begin{cases}
\binom{N_{\textrm{S}}-1}{ N_{\textrm{SI}}/2 -1} \binom{2}{2-d_{\textrm{I}}} \binom{N_{\textrm{I}}-2}{N_{\textrm{SI}}/2-(2-d_{\textrm{I}})} \quad & 2 \leq N_{\textrm{I}} \leq N-1 \\
1 \quad & N_{\textrm{I}} = 1 \quad \text{or} \quad N_{\textrm{I}}=N,
\end{cases}
\label{eq:A2.70}
\end{align}

\noindent
for all realisable values of $d_{\textrm{I}}$ and $N_{\textrm{SI}}$.

\end{appendices}


\end{document}